\documentclass[aps,prb,twocolumn,showpacs,superscriptaddress,groupedaddress]{revtex4}  
\usepackage{graphicx}  
\usepackage{dcolumn}   
\usepackage{bm}        
\usepackage{amssymb}   

\begin{document}


\title{Optimizing atomic resolution of force microscopy in ambient conditions}

\date{\today}

\author{Daniel S. Wastl}
\author{Alfred J. Weymouth} 
\author{Franz J. Giessibl} 

\affiliation{Institute of Experimental and Applied Physics, University of Regensburg,
Universit\"{a}tsstrasse 31, 93053 Regensburg, Germany}

\begin{abstract}
Ambient operation poses a challenge to AFM because in contrast to operation in vacuum or liquid environments, the cantilever dynamics change dramatically from oscillating in air to oscillating in a hydration  layer when probing the sample.
We demonstrate atomic resolution by imaging of the KBr(001) surface in ambient conditions by frequency-modulation atomic force microscopy with a cantilever based on a quartz tuning fork (qPlus sensor) and analyze both long- and short-range contributions to the damping. The thickness of the hydration layer increases with relative humidity, thus varying humidity enables us to study the influence of the hydration layer thickness on cantilever damping. Starting with measurements of damping versus amplitude, we analyzed the signal and the noise characteristics at the atomic scale. We then determined the  optimal amplitude which enabled us to acquire high-quality atomically resolved images.

\end{abstract}

\maketitle


\section{Introduction}

Today, atomic force microscopy (AFM) \cite{Binnig1986a} in frequency-modulation mode (FM-AFM) \cite{Albrecht1991a} allows us to routinely achieve atomic resolution in ultra high vacuum (UHV) \cite{Garcia2002,Morita2002bk,Giessibl2003,Morita2009bk}. For applications in  chemistry and biology at the nanoscale, high resolution research tools are needed for non-conductive and soft organic materials \cite{Kodera2010}.
The great advantage of FM-AFM over scanning tunneling microscopy \cite{BINNIG1983} is the ability to scan non-conducting surfaces  with true atomic resolution \cite{Bammerlin1998}. For biological as well as chemical samples, imaging  in their natural environment is often desired, requiring  AFM operation in air or liquid at room temperature (e.g. live cells, electrochemical studies, atomic study of chemical reactions and catalysis \cite{Rasmussen1998a}).
 High-resolution experiments are usually carried out in controlled environments like UHV at low temperatures to prevent the influence of drift, surface mobility of adsorbates or interaction with unwanted adsorbents. Ambient environments, where the surfaces under study are exposed to a mixture of gases and vapors, pose a profound challenge to surface studies requiring atomic resolution. The influences on the experiment are hard to predict in most cases, and while resolution down to the atomic scale was demonstrated to be feasible \cite{Wutscher2011}, it was not demonstrated until now in ambient conditions.\\

Obtaining atomic resolution in ambient conditions and liquids has proven to be more difficult than in UHV for two main reasons. First, in UHV, well defined surfaces can be prepared that stay clean for long times, while in ambient conditions, adsorbing and desorbing atoms and molecules can cause a perpetual change of the atomic surface structure on a time scale much faster than the time resolution of scanning probe microscopes. Second, the damping effects on the cantilever are quite well defined in UHV, where the quality factor $Q$ of the cantilever is high and often does not change significantly when bringing the tip close to the surface. For cantilevers oscillating in a liquid, the $Q$ factor is low but varies little with distance to the surface \cite{Labuda2013}. In contrast, for operation in ambient conditions, the $Q$ factor changes dramatically as the oscillating cantilever comes close the sample. Therefore, the excitation signal that is needed to drive the cantilever at a constant amplitude must increase profoundly, often by orders of magnitude, when approaching the oscillating tip towards the sample as it penetrates an adsorption layer. Surfaces in ambient conditions are usually covered by a water layer with a thickness that depends strongly on the relative humidity (RH) \cite{Israelachvili1991}. Correspondingly, the excitation amplitude has to increase when the oscillating force sensor moves from air into the water layer (this finding will be discussed in Fig.~\ref{water} and related text).\\

Atomic resolution  in liquid was obtained using quasi static AFM by Ohnesorge and Binnig \cite{Ohnesorge1993} in 1993, and recently Fukuma at al. obtained atomic resolution of the Muscovite mica surface in water using FM-AFM with standard micro fabricated cantilevers \cite{Fukuma2005b}. It has since been demonstrated by other groups on the calcite $(14\bar{1}0)$ cleavage plane in 1M KCl solution \cite{Rhode2009a} and  again on mica in water \cite{Hoogenboom2006}. Atomic resolution in liquid conditions using the qPlus sensor was demonstrated by Ichii et al\cite{Ichii2012}. \\

In this work, we analyze the dynamics of  FM-AFM measurements on the insulating and soft KBr(001) surface under ambient conditions with various tip materials and qPlus sensors. In section \ref{experimentalsetup} we describe the experimental  setup, where section \ref{allgemeinDempfung} starts with a description of cantilever motion in FM-AFM as a damped harmonic oscillator and presents the differences between working in UHV, liquid and ambient conditions. Section \ref{sample} introduces the potassium bromide sample and atomically resolved images on the (001) cleavage plane with different sensors. Section \ref{Ambientconditions} describes the ambient environment and several  effects of the adsorbed hydration layers including step movement (section \ref{setpmove}) and damping. Here the change in the damping  is clearly shown  to relate to the liquid layer height. In section \ref{SNRunderstand}, we  analyze  the signal (section \ref{normalizedSignal}) and noise (section \ref{noiseC}) in the hydration layer and determine the optimized  signal-to-noise ratio (SNR, section \ref{SNRanal}). To demonstrate this method, we discuss a fully worked example with both amplitude dependence and frequency shift dependence.\\

\section{\label{experimentalsetup}Experimental Setup}
\subsection{\label{allgemeinDempfung} FM-AFM cantilever, a damped harmonic oscillator}

The cantilever in FM-AFM is a damped driven harmonic oscillator \cite{Albrecht1991a}. The cantilever consists of a beam characterized by a stiffness $k$ and a resonant frequency $f_{0}$, with a sharp tip at the end. The tip oscillates at an amplitude $A$ such that the peak-to-peak distance is $2A$. Interaction with an external force gradient causes a frequency shift $\Delta f= f- f_0$ which is the observable in this operation mode. The frequency shift is related the force gradient  by $\Delta f(z)= -\frac{f_0}{2 k} \left\langle k_{\mathrm{ts}}(z)\right\rangle$, where $\left\langle k_{\mathrm{ts}}(z)\right\rangle$ is the averaged force gradient. In FM-AFM, an oscillation control circuit keeps the oscillation amplitude, and thus the  energy of the oscillation\cite{Giessibl2003} $E_{\mathrm{osc}}=\frac{1}{2}kA^2$ constant. This requires compensating both for internal dissipation (including friction in air) $\Delta E_{\mathrm{int}}$ and losses due to the tip-sample interaction (including friction in the water layer), $\Delta E_{\mathrm{ts}}$.\\
The losses or  damping in an oscillating  system can be described by the total energy loss $\Delta E_{\mathrm{tot}}= 2 \pi E_{\mathrm{osc}} / Q_{\mathrm{eff}}$  per oscillation cycle \cite{Giessibl2003}, where $Q_{\mathrm{eff}}$ is the effective quality factor.
Similarly, we can define $Q_{\mathrm{ts}}$ and $Q_{\mathrm{int}}$ by $\Delta E_{\mathrm{ts}}= 2 \pi E_{\mathrm{osc}}/Q_{\mathrm{ts}}$ and $\Delta E_{\mathrm{int}}= 2 \pi E_{\mathrm{osc}}/Q_{\mathrm{int}}$.
The quality factor in vacuum,  $Q_{\mathrm{vacuum}}$, or in air, $Q_{\mathrm{air}}$, can be determined by measuring a thermal oscillation spectrum of the free cantilever as described by Welker et al. \cite{Welker} or Giessibl et al. \cite{Giessibl2000, Giessibl2011a}. When the cantilever is solely driven by thermal energy, the equipartition theorem states that $E_{\mathrm{th}}=\frac{1}{2}k_{\mathrm{B}}T$, where each degree of freedom (kinetic and potential energy) of the cantilever holds a time-averaged energy of $E_{\mathrm{th}}$.

To maintain a constant oscillation amplitude greater than the thermal excitation, the sensor is driven by an external source that compensates for $\Delta E_{\mathrm{tot}}=\Delta E_{\mathrm{int}}+\Delta E_{\mathrm{ts}}$.
The signal that is needed to excite the beam is called the drive- or excitation-signal $V_{\mathrm{drive}}$ that causes a drive amplitude $A_{\mathrm{drive}}$, serving as a fingerprint for the energy losses. When the beam is excited at its resonance frequency, it oscillates at an amplitude $A=Q_{\mathrm{eff}} A_{\mathrm{drive}}$. An amplitude feedback circuit adjusts $A_{\mathrm{drive}}$ such that $A$ remains constant, thus a record of $Q_{\mathrm{eff}}$ as a function of sample position and distance can be deduced from $A_{\mathrm{drive}}$.
The relation between  $A_{\mathrm{drive}}$ and tip-sample dissipation $\Delta E_{\mathrm{tot}}$ has been shown to be \cite{Anczykowski1999}
\begin{equation}
\Delta E_{\mathrm{tot}}=\frac{\pi k A^2}{Q_{\mathrm{0}}}\left(\frac{A_{\mathrm{drive}}}{A_{\mathrm{drive,0}}}-1\right),
\label{eq:energyloss}
\end{equation}
where $A_{\mathrm{drive}}$ is the drive amplitude, $A_{\mathrm{drive,0}}$ is the drive amplitude far from the sample and $Q_{\mathrm{0}}$ is the quality factor far from the sample.
When additional energy losses $\Delta E_{\mathrm{ts}}$ occur during each oscillation cycle, the amplitude control circuit adjusts its drive voltage $V_{\mathrm{drive }}$ such that $A_{\mathrm{drive }}$ becomes greater than $A_{\mathrm{drive,0}}$, leading to a quality factor $Q_{\mathrm{ts}}$. The ratio between $A_{\mathrm{drive}}$ and $V_{\mathrm{drive}}$ is given by the sensitivity of the drive piezo, which is approximately 100\,pm/V in our setup.
\\

Eq. \ref{eq:energyloss} is  valid for a cantilever undergoing harmonic oscillation, i.e. when all the forces acting onto the cantilever can be treated as a small perturbation. This condition is met in our experiment due to the high stiffness of the qPlus sensors \cite{Giessibl2004}. We monitor the deflection of the cantilever in an oscilloscope and have also monitored higher harmonic components with an FFT spectrometer, showing that higher harmonics stay below the noise limit of about 1\,pm. At the same time, we monitor the magnitude of the fundamental amplitude with time, and find only slight variations in the range of one percent or so. \\

\begin{figure}
\includegraphics{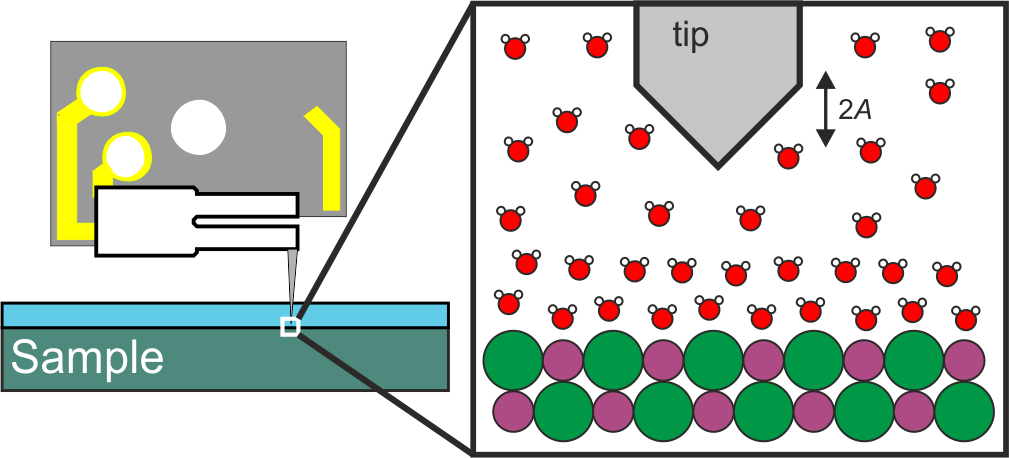}
\caption{\label{Qdepp} (Color online) Macroscopic picture of the environment when scanning under ambient conditions. Air, liquid and sample interaction contribute to cantilever damping. The blue layer on the sample represents the hydration layer, as described in the text (typically $<20\,$nm).   (b) Microscopic view when the tip is close to the surface. Here the bulk water and ordered hydration layers  on the sample are shown. We note that water layers on ionic crystals might be thicker than water layers on insoluble surfaces due to a screening of the ions by water molecules.}
\end{figure}

The full damping per oscillation cycle can be expressed with an effective quality factor  $Q_{\mathrm{eff}}$:
\begin{equation}
Q_{\mathrm{eff}}=\frac{1}{\frac{1}{Q_{\mathrm{int}}}+\frac{1}{Q_{\mathrm{ts}}}}
 \label{eq:Qeff}
\end{equation}
Driving a cantilever in vacuum only requires compensating for $\Delta E_{\mathrm{int}}$, because $\Delta E_{\mathrm{ts}}=0$. In air, $\Delta E_{\mathrm{ts}}$ is the damping of the cantilever due to interactions with air, resulting in a $Q_{\mathrm{air}} < Q_{\mathrm{vacuum}}$.
In ambient conditions, a sample can be covered by a water layer.  This causes an additional increase of $\Delta E_{\mathrm{ts}}$ which is highly dependent up on the height of the hydration layer and its molecular structure close to the sample. As $\Delta f$ is related to the force gradient, the conservative forces at play can be evaluated. The drive signal gives access to the dissipative forces.
 \\
We use qPlus sensors which are self-sensing and based on quartz tuning forks \cite{Giessibl1998,Giessibl2000,Giessibl2011a}. The qPlus sensor was originally used in ambient environments \cite{Giessibl1998}, and since the year 2000 when we obtained atomic resolution in UHV \cite{Giessibl2000}, we tried to achieve atomic resolution in ambient conditions as well. An unique feature of our setup is the length of the bulk tips we use. By approaching the sensor with a tip length $l_{\mathrm{tip}}\approx 500\, \mu$m most of the tip remains outside the hydration layers.
In Fig.~\ref{Qdepp} this is shown from both a macroscopic and microscopic perspective. In Fig.~\ref{Qdepp}~(b) ordered hydration layers are shown on the atomically flat sample surface. This situation will be discussed in depth  in section \ref{effectondamping}.
Another key improvement was  the use of a digital amplitude controller (OC 4  from Nanonis/SPECS, CH-8005 Zurich, Switzerland) that has a very large dynamic range and is able to adjust the excitation signal by orders of magnitude. The microscope head was a UHV-compatible microscope with a double-stage spring suspension system\cite{Giessibl1994a} used in ambient air. For operation in very low humidity, the microscope can be bolted onto a small metal can containing a bag of silica gel. \\

\subsection{\label{sample}Sample: potassium bromide}
\begin{figure}
 \includegraphics[scale=1.0]{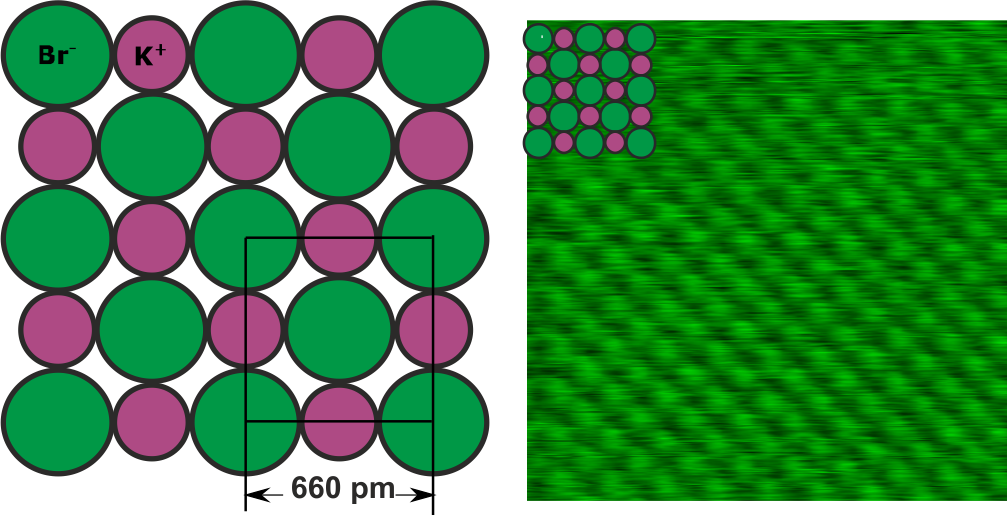}
 \caption{(Color online) (a) KBr(001) cleavage plane, with a lattice constant of $a_0= 660\,$pm. Bare ionic radius  is $r_{\text{Br}}=195\; \text{p}\text{m}$ for the $\text{Br}^{-}$ ions (blue) and  $r_{\text{K}}= 133 \;\text{p}\text{m}$ for the $\text{K}^{+}$ ions (white) \cite{Giessibl1992b}. (b) Topographic image with atomic resolution of the KBr(001) cleavage plane. Operating parameters $\Delta f=  +190\,$~Hz, $A= 75 \; \text{pm}$, $f_{0}=38853\,$~Hz and $k =1000\,\frac{\text{N}}{\text{m}}$, bulk sapphire tip.    \label{Fig01}}
 \end{figure}
\begin{figure}
 \includegraphics[scale=1.0]{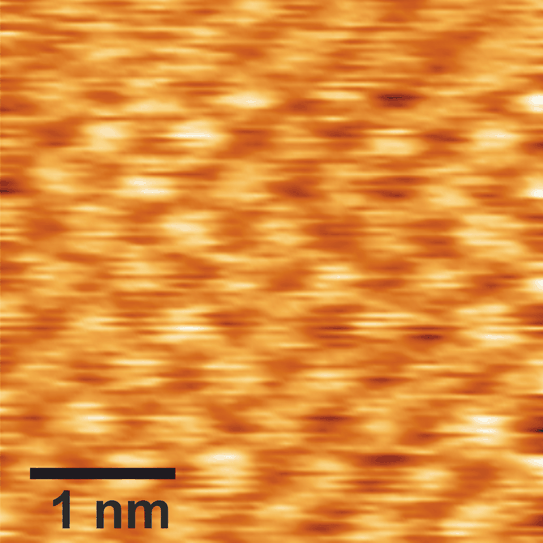}
\caption{(Color online) Topographic image with atomic resolution of the KBr(001) cleavage plane (lattice constant $660 \; \text{pm} $). Flat terrace imaged at $\Delta f=  +205\,$~Hz and an oscillation amplitude of $A= 60 \; \text{pm}$. Sensor parameters: $f_{0}=20719\,$~Hz and $k =1280\,\frac{\text{N}}{\text{m}}$ with an etched bulk tungsten tip.    \label{Fig02}}
\end{figure}

Over the last decades, insulators in the form of ionic crystals have been studied by atomic force microscopy in vacuum at both room temperature \cite{Meyer1990,Meyer1990a} and low temperature \cite{Giessibl1992b}. Atomic resolution on terraces and steps has been reported on bulk ionic crystals in UHV \cite{Bammerlin1998,Bennewitz2002,Maier2007,Pawlak2012}.
To date, insulators like sodium chloride or potassium bromide are used for basic research both in bulk crystalline form and as thin films serving as spacing layers on metal surfaces \cite{Repp2004, Gross2009}. These applications include, e.g., imaging of individual molecule orbitals \cite{Gross2009,Repp2005} and molecular switches \cite{Pavlicek2012}, organic structure determination \cite{Gross2010} and investigations of friction on the nanoscale \cite{Filleter2008}.\\
Potassium bromide (CrysTec Kristalltechnologie, D-12555 Berlin, Germany) crystals were prepared by cleaving in air with a blade along the (001) plane.
Potassium bromide  has a NaCl structure with a lattice constant of $a_0=660 \;\text{pm}$ (see Fig.~\ref{Fig01}~(a)). The bare ionic radius  is $r_{\text{Br}}=195\; \text{p}\text{m}$ for the $\text{Br}^{-}$ ions and  $r_{\text{K}}= 133 \;\text{p}\text{m}$ for the $\text{K}^{+}$ ions \cite{Giessibl1992b}.
 Following earlier publications, only the large $\text{Br}^{-}$ ions should be visible \cite{Meyer1990a}.
Figs.~\ref{Fig01}~(b) and \ref{Fig02} shows flattened data of  atomically resolved images taken on a freshly cleaved KBr crystal in air, imaged with a bulk sapphire tip (Fig.~\ref{Fig01}~(b)) and etched tungsten tip (Fig.~\ref{Fig02}). The scans shows the ionic structure of the KBr(001) cleavage plane with a lattice constant of $660 \; \text{pm}$. The square lattice with a spacing of $460 \; \text{pm} $ corresponds well to the spacing between equally charged ions of $\frac{\text{a}_0}{\sqrt{2}}$. Therefore this square lattice represents the unreconstructed $1\; \times \; 1$ surface structure, exposing only one atomic species \cite{Meyer1990a,Meyer1990} (presumably the one with the greater ionic radius, here Br).
 We assume that surface material is attached to the tip apex during collisions with the surface, which creates a polar tip that facilitates atomic resolution on ionic surfaces \cite{Hofer2003}.
One striking feature in these images is that no defects can be seen. This can be explained by the surface being covered with a hydration layer that is likely to consist of a saturated solution of $\text{K}^{+}$ and  $\text{Br}^{-}$ ions in $\text{H}_{2}\text{O}$. Even if atomic defects are created by thermal excitation, they exist for much shorter time spans than the time span accessible to our force microscope.\

\section{\label{Ambientconditions} Ambient conditions: the hydration layer}

The term \lq\lq ambient conditions\rq\rq{} refers to a poorly defined state that involves a large number of variable parameters; the laboratory air mainly consists of oxygen, nitrogen, carbon-oxides and rare-gases. The individual concentration of these gases is usually not controlled, but has an effect on the sample, e.g. oxidation. Laboratory air also has a significant amount of water vapor, where the partial pressure of water depends on temperature and relative humidity (RH). If RH is greater than zero, all surfaces (hydrophobic or hydrophilic) exposed to air \cite{Palasantzas2009,James2011} adsorb a water-layer with a thickness dependent on the exposure time, temperature, RH and the sample's hydrophilic or hydrophobic character\cite{Davy1998, Wei2000, Huang2007, Freund1999}.

\subsection{\label{setpmove} Clear indication of the liquid layer: step motion}

\begin{figure}
\includegraphics[scale=1.0]{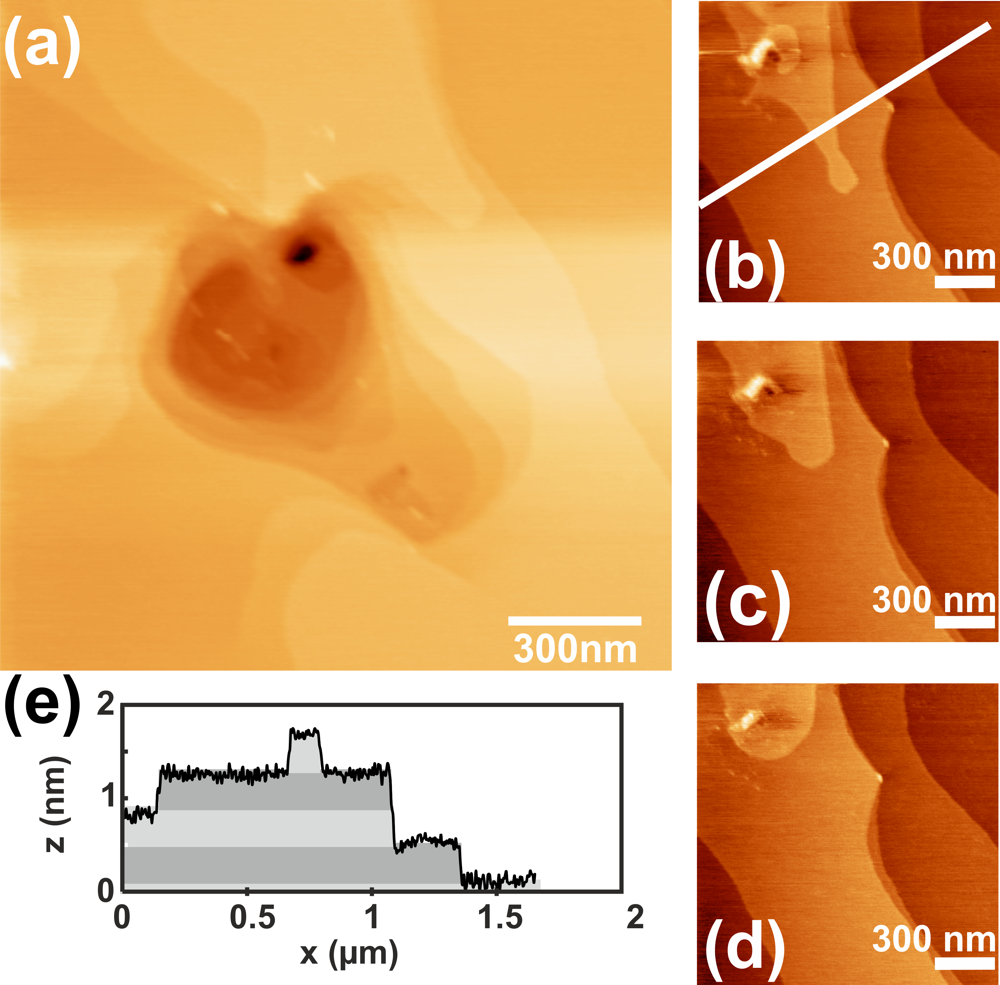}
\caption{\label{move}(Color online)(a) Sub-micron defect, created by indenting an iron \cite{Schneiderbauer2012} tip into the KBr(001) surface. The defect consists of a hole with a diameter of approx. 300\,nm and a depth of about 5 nm, surrounded by mono-atomic terraces. The image is acquired using the same tip that created the indentation. Scan-parameters: $\Delta f = +50 \; \text{Hz}$, $A = 400 \; \text{pm}$, $f_{0}=26691 \text{Hz}$,  $k =1280\,\frac{\text{N}}{\text{m}}$, bulk iron tip, RH$\, \approx \, 60\%$. (b) to (d): Time evolution of the AFM image taken at  (b) $t_{0} = 0\;\text{s}$, (c) $t_{1} = 1140\;\text{s}$ and (d) $t_{2} = 4200\;\text{s}$, showing the dissolution of the top terrace with time. (e) Line profile of (b) indicating single and double atomic steps.}
\end{figure}

A clear indication of the presence of a liquid-layer with dissolved K$^+$ and Br$^-$ ions on the surface is the rapid movement of steps shown in Fig.~\ref{move}(b)-(d). Filleter et al. showed that by poking the surface, the tip could be used to create mono-atomic terraces due to plastic deformation of the KBr(001) cleavage plane in UHV \cite{Filleter2006}. Following the UHV experiments we used this nanoindentation method for both tip preparation and  to create steps on the flat KBr(001), by poking the tip approximately $100\,$~nm into the surface. In Fig.~\ref{move}, the surface can be seen after a nanoindentation experiment. 	
The indentation of a tip is surround by terraces and steps, similar to those reported in the UHV experiments \cite{Filleter2006}. In Fig.~\ref{move}~(b) and the corresponding line scan from Fig.~ \ref{move}~(e), mono- and di-atomic steps can be seen with an height of $\approx 330\,$~pm and $660\,$~pm.
We found that steps, created by the nanoindentation on the KBr(001) surface (Figs.~\ref{move}~(b) to (d)), dissolve rapidly with time. This is similar to previous investigations \cite{Luna1998}, where step motion on KBr as a function of RH was reported. The time delay between Figs.~\ref{move}~(b) and \ref{move}~(d) of $\Delta t = 4200\;\text{s}$ demonstrates that the steps are not only mobile directly after poking. From this data we extract a speed of step motion of approximately $100 \, {\text{pm}}/{\text{s}}$, which is too fast to image with atomic resolution. \\
The movement is also present at naturally occurring steps. At room temperature, steps are mobile due to adsorption/desorption of $\text{K}^{+}$ and $\text{Br}^{-}$ ions that are readily available from the hydration layer (saturated solution of $\text{K}^{+}$ and $\text{Br}^{-}$ ions in water).
However, as we show in the following, the key challenge in ambient operation is the strong variation of the cantilever damping (and $Q$) as a function of distance and of the oscillation amplitude near the sample due to the adsorption layer.

\subsection{\label{effectondamping} Effect of hydration layer on damping}
\begin{figure}
\includegraphics[scale=1.0]{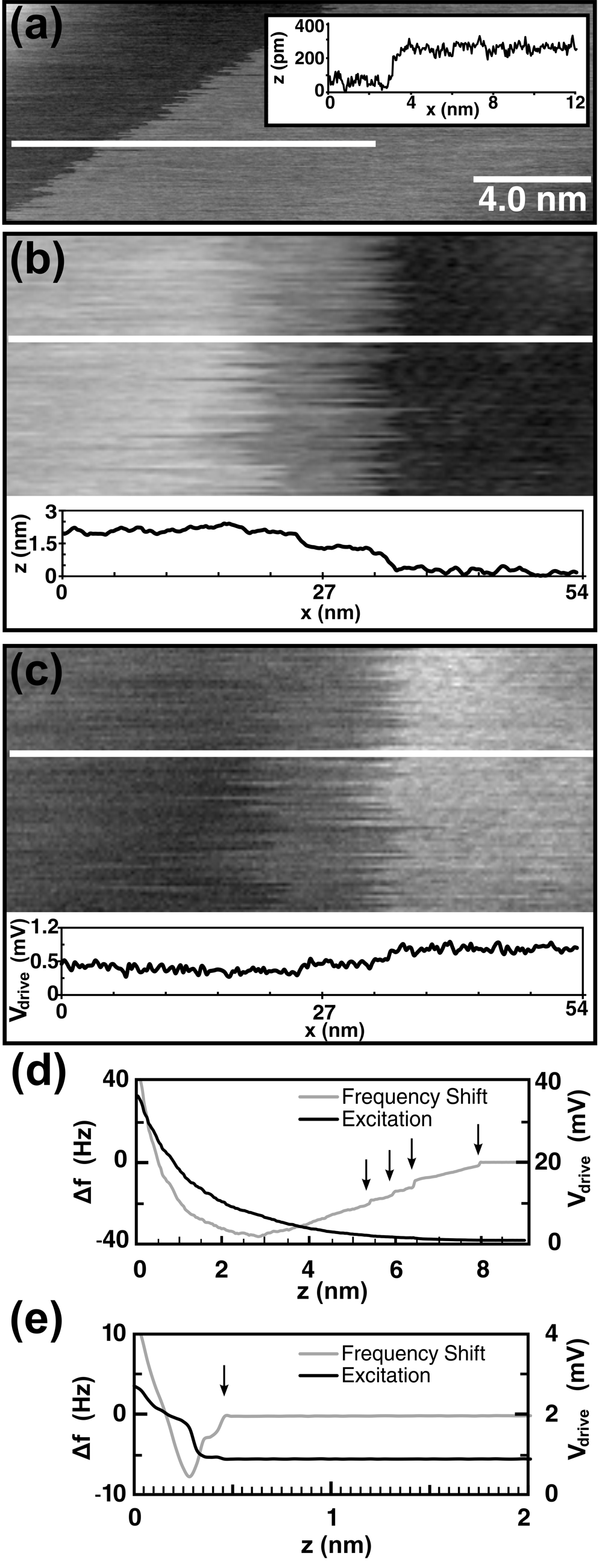}
\caption{\label{water} (a) Topographic  image of a water mono-layer on cleaved KBr(001) left three days in air at RH $\, \approx \, 53$\%.  $\Delta f=+9.6$\,Hz,  $A = 780\,$~pm,  $f_{0}=24071$~Hz and $k =1800\,\frac{\text{N}}{\text{m}}$ with a bulk sapphire tip.   (b) Topographic image of a cleaved KBr(001) crystal after exposure to air at RH $\, \approx 60$\% for two hours and (c) excitation signal.  $\Delta f=+16.6$\,Hz,  $A = 104$\,~pm,  $f_{0}=31464$ Hz and $k =1800\,\frac{\text{N}}{\text{m}}$ with an bulk sapphire tip.
(d)  $\Delta f(z)$ and  $V_{\mathrm{drive}}(z)$ of the sample shown in (a)
(e)  $\Delta f(z)$ and  $V_{\mathrm{drive}}(z)$ taken after drying the sample with a heat-gun.}
\end{figure}

In Fig. \ref{water} (a) we show step-like structures on the KBr crystal which was scanned with a sapphire tip in air in a RH of 60\%.
While these step edges change much slower with time than the mono-atomic steps shown in Fig. \ref{move} (b)-(d), they are not stationary  and the height of a single layer step which is shown in Fig. \ref{water} (a) is approximately $200\,$~pm (see the inset linescan in Fig.~\ref{water}~(a)), notably less than the $330\,$~pm height of KBr-mono-steps shown in Fig.~\ref{move}~(e) the line scan of  Fig.~\ref{move}~(b) (low-pass filtered).
Our hypothesis for the origin of the steps in Fig.~\ref{water}~(a) and (b) is they are due to single and multiple additional water layers on the KBr crystal, because these step heights are close to the thickness of a single hydration layer \cite{Jeffery2004,Fukuma2007,Kimura2010,Israelachvili1983,Kilpatrick2013}. Figure~\ref{water}~(c) shows that more energy is required to maintain a constant drive signal when penetrating the water layers. As more water layers are penetrated by the tip, a larger excitation signal is required  which is clearly visible in the steps in the inset in Fig.~\ref{water}~(c). It is interesting to note the relatively sharp edges where the water layer is penetrated. We speculate that this edge is either due to a domain boundary of the water layer or a
possible Moire effect, where the sticking of the water layer to the ionic crystal varies laterally due to a lattice mismatch.  \\

 On the lower terrace in Fig.~\ref{water}~(b)  a molecular ordered structure appears, possibly due to ``icelike'' water on the KBr(001) surface. The existence of an ``icelike'' water mono-layer on mica at room temperature  was reported by Miranda \cite{Miranda1998}. While no structural information was given, there are indications of  ordering of the water molecules on the surface. In our data, we resolve a periodicity of 1.9 nm. Experimental \cite{Foelsch1992} and theoretical studies\cite{Wassermann1993, Meyer2001, Park2004} elucidated the structure of adsorbed water on (100) cleavage planes of related alkali halide surfaces. LEED Experiments by F{\"o}lsch et. al. \cite{Foelsch1992} on a NaCl(100) substrate showed a well-ordered ice-like c$(2\times4)$ bilayer structure of water molecules. This structure is similar to that of ordinary  $\text{I}_h$ ice, except that the adsorbed bilayer is slightly distorted due to the lattice mismatch with the NaCl(100) surface \cite{Peters1997}. The experimental findings of the well-ordered ice-like c$(2\times4)$ bilayer structure are supported by theoretical approaches including molecular dynamics calculations by Wassermann  et al.\cite{Wassermann1993} and Meyer et al.\cite{Meyer2001} as well as density functional calculations from Park et al. \cite{Park2004} \\

Considering that the binding energy of $\text{H}_{2}\text{O}$ molecules to an alkali halide  crystal is on the order of 0.4~eV \cite{Ewing2006}, it is reasonable to observe higher dissipation on sample areas where this water layer is expelled from the surface. Hydrodynamic friction forces could also contribute to energy dissipation that occur when the water layer that separates tip an sample is expelled and drawn in by the oscillating tip \cite{Labuda2013}.
A study of frequency shift versus distance and excitation versus distance  gives further insight into the effect of the water layers on imaging. Fig.~\ref{water}~(d) shows a spectrum taken in typical ambient conditions, that is, with a RH of approximately 53\%.
Jumps in the $\Delta f(z)$ signal are indicated by arrows. We propose two possible explanations: i) the breaking of the hydration layers or ii) a molecular-scale rearrangement in the water meniscus as the tip retracts from the surface. The use of small oscillation amplitudes less than 1\,nm is helping to observe these fine details.\\

The total measurable interaction extends nanometers from the surface.  More importantly, the excitation signal $V_{\mathrm{drive}}(z)$ increases from $0.5\,$~mV  far from the surface to $40\,$~mV near the surface. In order to test the water film hypothesis, we dried the sample by heating with a heat-gun and quickly acquired a $\Delta f(z)$- and $V_{\mathrm{drive}}(z)$- spectrum thereafter. While we could once again resolve atomic contrast, both $\Delta f(z)$ and $V_{\mathrm{drive}}(z)$, shown in Fig.~\ref{water}~(e), are drastically different. Now the excitation near the surface is $<\, 5\,$mV and there is evidence of only one water layer.

The frequency shift and damping spectra in Fig.~\ref{water}~(d) and (e) highlight one of the profound challenges for observing atomic resolution in ambient conditions. The increase of damping as the tip  penetrates the liquid layer poses challenges for the amplitude controller in maintaining a constant amplitude. In typical conditions the humidity is so large that several water layers form on any surface \cite{Palasantzas2009,James2011}. The effect of this is a drastic lowering of $Q$ and the dramatic increase of $\Delta E_{\mathrm{ts}}$ where the tip is close enough to the surface to resolve atoms. This can be seen by the large excitation required in Fig.~\ref{water}~(d).

\begin{figure}
\includegraphics[scale=0.9]{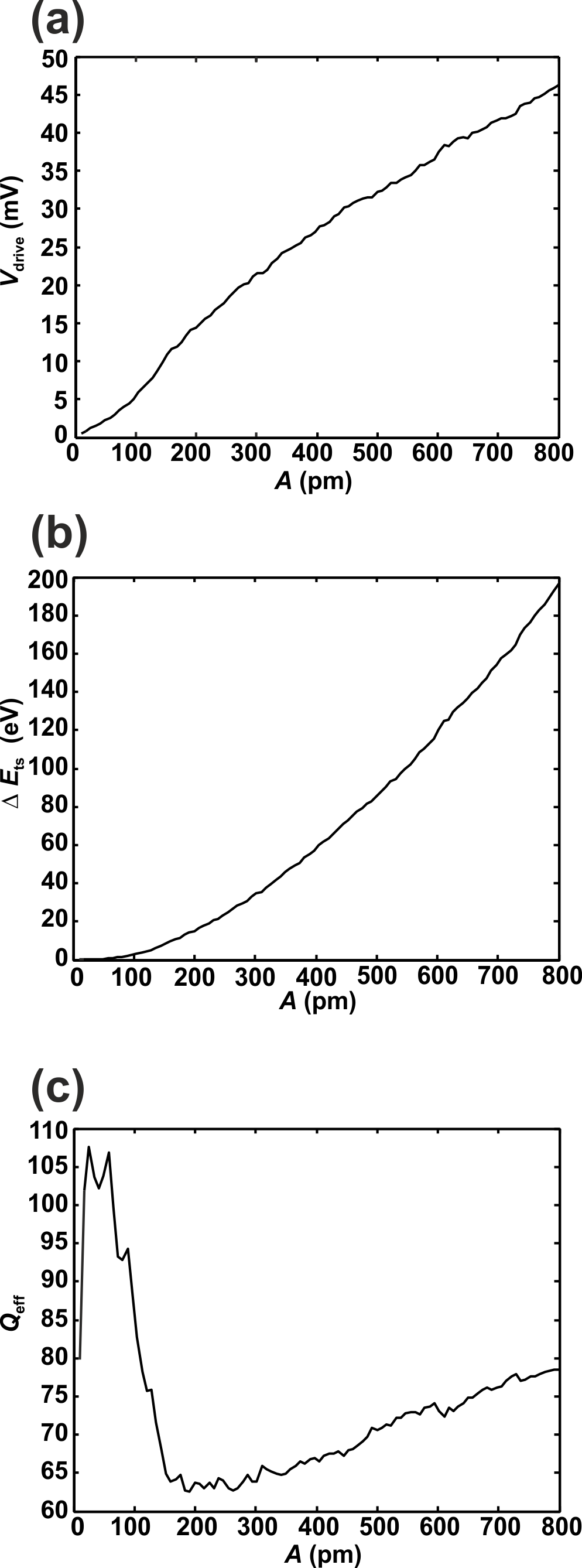}
\caption{\label{DriveENERGIEQ} (a) Excitation, (b) Energy loss and (c) quality factor as a function of amplitude.  Acquired with a qPlus sensor with an etched tungsten tip.  $f_{0}=23321\,$Hz $k=1280 \,$ $\frac{\text{N}}{\text{m}}$, $Q_{\mathrm{air}}=2000$, $\Delta f=190\,$Hz.}
\end{figure}

To further investigate the effect of the hydration layer we record the excitation signal $V_{\mathrm{drive}}$ while the sample is in intermittent contact with the tip for a constant frequency shift as a function of the oscillation amplitude.
Figure~\ref{DriveENERGIEQ}~(a) covers a variation of the oscillation amplitude $A$ from $10\,$pm to $800\,$pm at a constant frequency shift of $\Delta f = 190\,$Hz, resulting in an increase of $V_{\mathrm{drive}}$ from about 1\,mV to 46\,mV.
Figure~\ref{DriveENERGIEQ}~(b) depicts the energy loss $\Delta E_{\mathrm{ts}}$  calculated with the Eq.~\ref{eq:energyloss}. Here, the dramatic increase of  energy loss in $\Delta E_{\mathrm{ts}}$ versus oscillation amplitude  $A$ is clearly visible.
The amplitude dependence of the effective damping factor $Q_{\mathrm{eff}}$ described by Eq.~\ref{eq:Qeff} is plotted in Fig.~\ref{DriveENERGIEQ}~(c), showing most interesting features for small amplitudes below $1\,$nm. Here, a steady decrease of the quality factor occurs until an oscillation amplitude of $A\approx 300\,$pm is reached, where a wide minimum in the range of $A\approx 300-150\,$pm occurs before $Q_{\mathrm{eff}}$ rises to a plateau for amplitudes around 60\,pm.
When assuming a water layer on the surface with a thickness of the first hydration  layer of  $\approx 200-310\,$pm \cite{Jeffery2004,Fukuma2007,Kimura2010,Israelachvili1983,Kilpatrick2013}, the tip would entirely remain within the ordered hydration layer if its amplitude is smaller than  $\approx 150\,$pm. At larger amplitudes, the first ordered hydration layer would be penetrated during each oscillation cycle. According to this notion, dissipation would be low for peak-to-peak amplitudes smaller than the thickness of one water layer, leading to a high $Q_{\mathrm{eff}}$. The amplitude dependence of $Q_{\mathrm{eff}}$ has implications on the noise of the AFM signal which we will discuss in the following section.

\section{\label{SNRunderstand} Signal-to-noise-ratio in ambient conditions}

\subsection{\label{normalizedSignal} The dependence of frequency shift (signal) with amplitude}
In FM-AFM  the frequency shift  $\Delta f$ is a measure of  the average force gradient $\left\langle k_{\mathrm{ts}}(z)\right\rangle$ as explained in section \ref{allgemeinDempfung}.
In order to model the  average force gradient $\left\langle k_{\mathrm{ts}}(z)\right\rangle$, we use an exponential force law $\propto \text{exp}(\frac{-z} {\lambda})$ with a decay constant $\lambda$ \cite{Giessibl1992a}. In the case of ionic crystals $\lambda$ has been shown to be $\lambda=\frac{a}{2 \pi}$   \cite{Giessibl1992a} where  $a=\frac{a_0}{\sqrt{2}}$. Using the  lattice constant $a_0$ for KBr, we get a $\lambda$ of 75\,pm. The signal is then the force gradient convolved over the tip oscillation, as shown in Ref\cite{Giessibl2003}:
\begin{equation}
\left\langle k_{\mathrm{ts}}(z)\right\rangle \propto \frac{2}{\pi A^2}\int_{-A}^{A}e^{-\frac{z+A-q}{\lambda}}\sqrt{A^2-q^2}dq \, ,
\end{equation}
where $z$ is the distance between sample and oscillating tip.
By integrating  and considering only the $z$ independent terms at a constant point of closest approach,  the normalized model signal is then:

\begin{equation}
S_{\mathrm{normalized}} \propto  \frac{2\lambda}{A} e^{-\frac{A}{\lambda}} I_{1}\left(\frac{A}{\lambda} \right)\, .
\label{eq:signal}
\end{equation}

The normalized signal is plotted in Fig.~\ref{SNR2} (dashed dotted-line) for the case of $\lambda=75\,$~pm  in an interval of $10\,$~pm to $800\,$~pm.
Scanning with stable  oscillation is possible down to oscillation amplitudes of $10\,$pm for our ambient qPlus setup, depending on the sensor.
One should notice that already  at an amplitude of $50\,$~pm the signal has  decreased to 55\% of the maximum. However, one has to also consider  the noise as a function of amplitude.

\subsection{\label{noiseC} The dependence of the effective quality factor $Q_{\mathrm{eff}}$ and noise with amplitude}

Three sources dominate noise in frequency modulation AFM:  thermal-, detector- and oscillator noise. These noise sources are small for low deflection detector noise densities $n_{\mathrm{q}}$ (the ratio between the electrical noise density and the sensitivity $S$ of the electrical signal \cite{Giessibl2011a}) and high $Q$ factors.
An in depth discussion of the noise terms and there origin is given in \cite{Albrecht1991a, Giessibl2011a, Wutscher2011}.\\
The   minimum detectable average force gradient $\delta \left\langle k_{\mathrm{ts}} \right\rangle_{min}$ is given by:
\begin{equation}
\delta \left\langle k_{\mathrm{ts}} \right\rangle_{\mathrm{min}}=\sqrt{\delta k_{\mathrm{ts,th}}^2 + \delta k_{\mathrm{ts,det}}^2+\delta k_{\mathrm{ts,osc}}^2}
\label{eq:Noise}
\end{equation}
The force gradients for the thermal-, detector-, and oscillator frequency noise are given by \cite{Wutscher2011}:

\begin{equation}
\delta k_{\mathrm{ts,th}}= \sqrt{\frac{4 k k_B T B}{\pi f_0 A^2 Q}} \propto \frac{1}{AQ^{\frac{1}{2}}}
\end{equation}

\begin{equation}
\delta k_{\mathrm{ts,det}}= \sqrt{\frac{8}{3}}\frac{k n_q B^{\frac{3}{2}}}{f_0 A} \propto \frac{1}{A}
\end{equation}

\begin{equation}
\delta k_{\mathrm{ts,osc}}= \frac{k n_q \sqrt{2B}}{ A Q} \propto \frac{1}{A Q}
\end{equation}
here $k$ is the stiffness, $f_0$ the resonance frequency, $T$ temperature, $B$ bandwidth, $A$ oscillation amplitude, $k_B$ Boltzmann constant.
The equations point out that all noise terms are proportional to $\frac{1}{A} $ and that $Q$ plays an important role in both thermal and oscillator noise.
Usually,  $Q$ is assumed to be constant with  oscillation amplitude. It has been calculated that a constant energy loss per oscillation cycle leads to an amplitude dependence of $Q$ \cite{Giessibl1999a}. Here, we have shown experimentally that the quality factor is amplitude dependent. We use the experimental dependence $Q(A)$ to calculate the amplitude-dependent noise explicitly, shown by the dotted line in Fig. \ref{SNR2}. This will be used in the following section to determine the amplitude for the optimal SNR, which leads to the optimal imaging parameters for atomic resolution.

\subsection{ \label{SNRanal}Signal-to-noise ratio}

\begin{figure}
\includegraphics[scale=1.0]{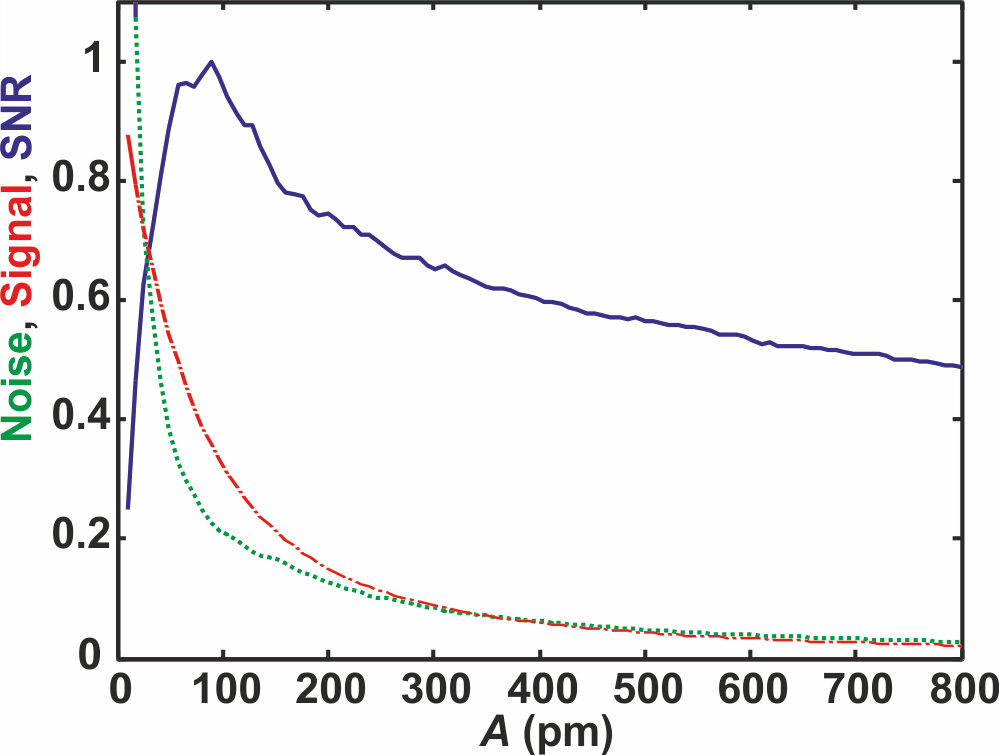}
\caption{\label{SNR2} (Color online) Diagram of signal (dashed dotted-line), noise (dotted-line) and normalized SNR (solid-line) vs amplitude, where the maximal SNR is arbitrarely set to 1. Data taken with a qPlus sensor ($k = 1280\,\frac{\text{N}}{\text{m}}$, $f_0=23321\,$ Hz, $Q_{\mathrm{air}}=2100$)  equipped with an etched tungsten tip at RH = 31\%. Signal was calculated with Eq.~\ref{eq:signal}, notice its maximum at $10\,$pm is 90\%. The noise was calculated by used of Eq.~\ref{eq:Noise} with the $Q_{\mathrm{eff}}$ values calculated due to the $A_{\mathrm{drive}}$ spectra taken at a frequency shift of $190\,$Hz.}
\end{figure}

With the amplitude dependence of $Q_{\mathrm{eff}}$, we can analyze the SNR for the data set of Fig.~\ref{DriveENERGIEQ}. Fig.~\ref{SNR2} shows the SNR graph (solid line) which is calculated from the  noise  with the $Q_{\mathrm{eff}}$ shown in Fig.~\ref{DriveENERGIEQ}~(c)  and the model signal calculated with Eq.~\ref{eq:signal}.
Fig.~\ref{SNR2} shows a large peak at low amplitudes where the best imaging amplitude with the highest contrasts is expected at $A_{\mathrm{opt}}=89\,$pm. \\
In the  following section we demonstrate the optimization of the scan parameters which lead to the best atomic contrast.

\begin{figure}
\includegraphics[scale=1]{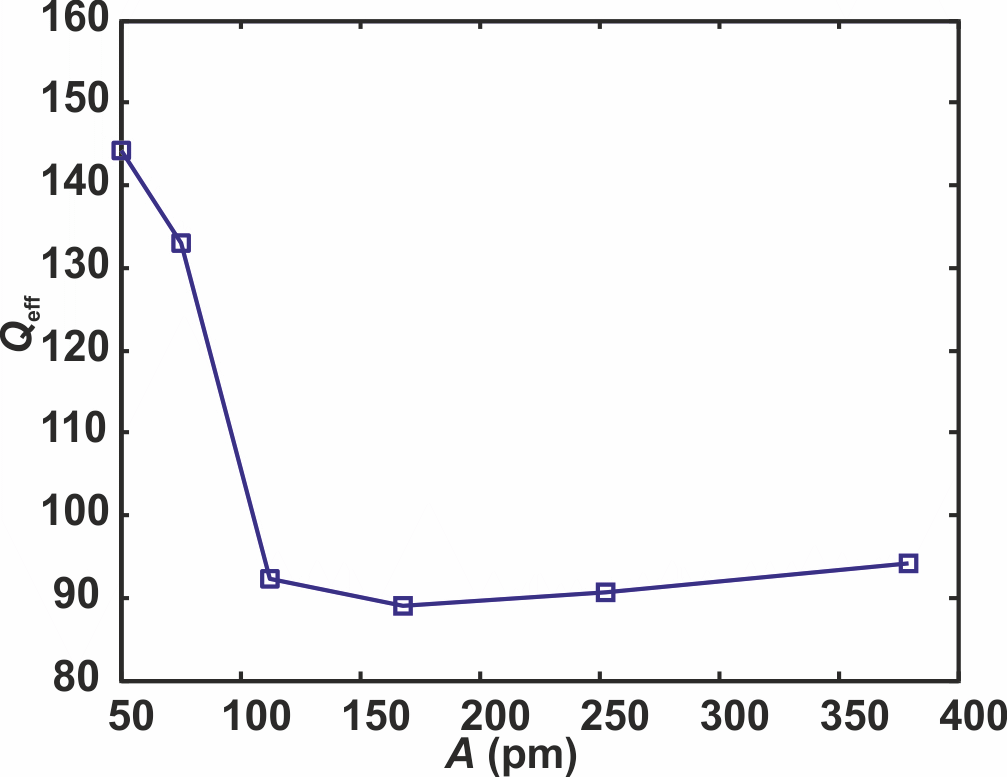}
\caption{\label{Qeffdiagram} (Color online) Diagram of $Q_{\mathrm{eff}}$ vs amplitude. The values are calculated from the excitation data corresponding to the images in Fig.~\ref{optimal} as  described in the text.}
\end{figure}

\section{\label{FWE} Signal-to-noise optimization for a specific experiment}

In the following a qPlus sensor with a stiffness of $k = 1000\,\frac{\text{N}}{\text{m}}$ and sapphire tip  (splinters from a bulk sapphire crystal) is used.  First we determine the free excitation signal $V_{\mathrm{drive,0}}$ in air versus the oscillation  amplitude $A$. After approaching on a freshly cleaved KBr(001) surface plane we begin the tip modification by poking. Nanoindented holes like those shown in Fig.~\ref{move}~(a) are the result of controlled pokes that modify the tip apex favorably to enable atomic resolution. After some time, tips appear stable in large scale images.  Using these stable tips, we measured the excitation signal $V_{\mathrm{drive}}$ as a function of amplitude while the closed $z$-feedback loop adjusted a constant frequency shift of $\Delta f =190\,$Hz. We used the method discussed in the previous section to calculate $Q_{\mathrm{eff}}$ from the excitation $V_{\mathrm{drive}}$.

\begin{figure}
\includegraphics[scale=1.0]{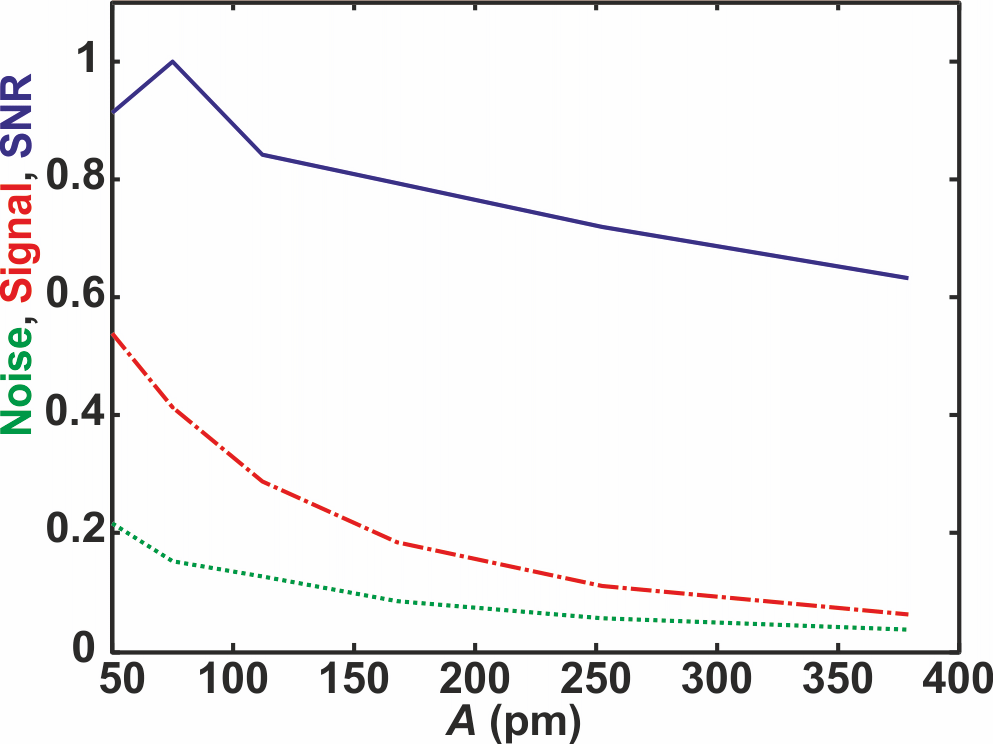}
\caption{\label{SNR} (Color online) Diagram of signal (dashed dotted-line), noise (dotted-line) and normalized SNR (solid-line) vs amplitude. Data taken with a qPlus sensor ($k = 1000\,\frac{\text{N}}{\text{m}}$, $f_0=38853\,$ Hz, $Q_{\mathrm{air}}=2977$) equipped with a bulk sapphire tip, at RH = 35\%. Signal was calculated with Eq.~\ref{eq:signal}. The noise was calculated by used of Eq.~\ref{eq:Noise} with the $Q_{\mathrm{eff}}$ values of the real data shown in Fig.~\ref{Qeffdiagram}.}
\end{figure}

Fig.~\ref{Qeffdiagram} shows the effective damping $Q_{\mathrm{eff}}$ as a function of amplitude.
Both Figs. \ref{DriveENERGIEQ} (c) and \ref{Qeffdiagram} have a similar shape and show the same key features, including a nearly stable but very low $Q_{\mathrm{eff}}$ for amplitudes $A$ greater than half the height of one hydration layer.\\
 Using the calculated  $Q_{\mathrm{eff}}$, we find the noise for this sensor as a function amplitude. We use Eq.~\ref{eq:signal} to calculate the normalized model signal, again using $\lambda=75\,$pm for KBr(001).
Figure~\ref{SNR} shows  the calculated noise  with the $Q_{\mathrm{eff}}$ shown in Fig.~\ref{Qeffdiagram} (dotted line), the normalized model signal calculated with Eq.~\ref{eq:signal} (dashed dotted line) and the SNR graph (solid line). The signal-to-noise-ratio is normalized to a maximum of one in the diagram.
From this graph, the optimal SNR occurs at an amplitude of $A_{\mathrm{opt}}=75\,$pm. The value of $A_{\mathrm{opt}}=75\,$pm for the sapphire tip is close to the value of the tungsten tip of $A_{\mathrm{opt}}=89\,$pm. 

\begin{figure*}
\includegraphics[scale=1.0]{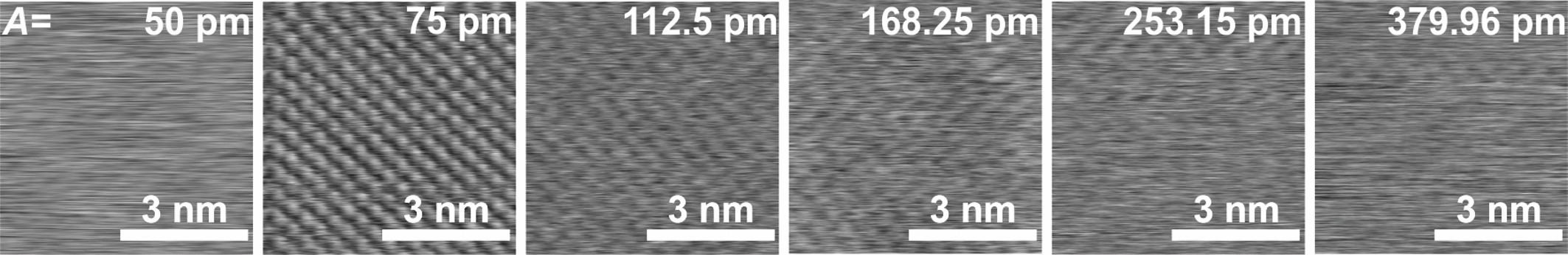}
\caption{\label{optimal} Amplitude dependence of the signal at   $\Delta f_{\mathrm{opt}}=190\,$Hz. Amplitudes are shown in each frame. Images are  shown in which  atomic contrast was able to be seen without being filtering.   Data taken with a qPlus sensor ($k = 1000\,\frac{\text{N}}{\text{m}}$, $f_0=38853\,$Hz)  equipped with a sapphire tip. Data were line flatted.}
\end{figure*}
Figure~\ref{optimal} shows a set of atomically resolved images at a constant set-point of $\Delta f =190\,$Hz, starting with an oscillation amplitude of $50\,$pm and consecutively increasing it by a factor of 1.5. These images were flattened but not filtered. The maximum amplitude where atomic resolution was obtained is $A=380\,$pm. The largest contrast was found for an oscillation amplitude of $A=75\,$pm. This is in good agreement with the calculated optimal SNR discussed above. \\
These observations strongly support our hypothesis that this is an ionic imaging mechanism with a decay constant of  $\lambda=75\,$pm.
Even without poking the tip  it is very likely that the tips are terminated by surface material of the sample due to scanning. Our hypothesis is that these light pokes only modify the tips front cluster, enabling atomic resolution.

\begin{figure*}
\includegraphics[scale=1.0]{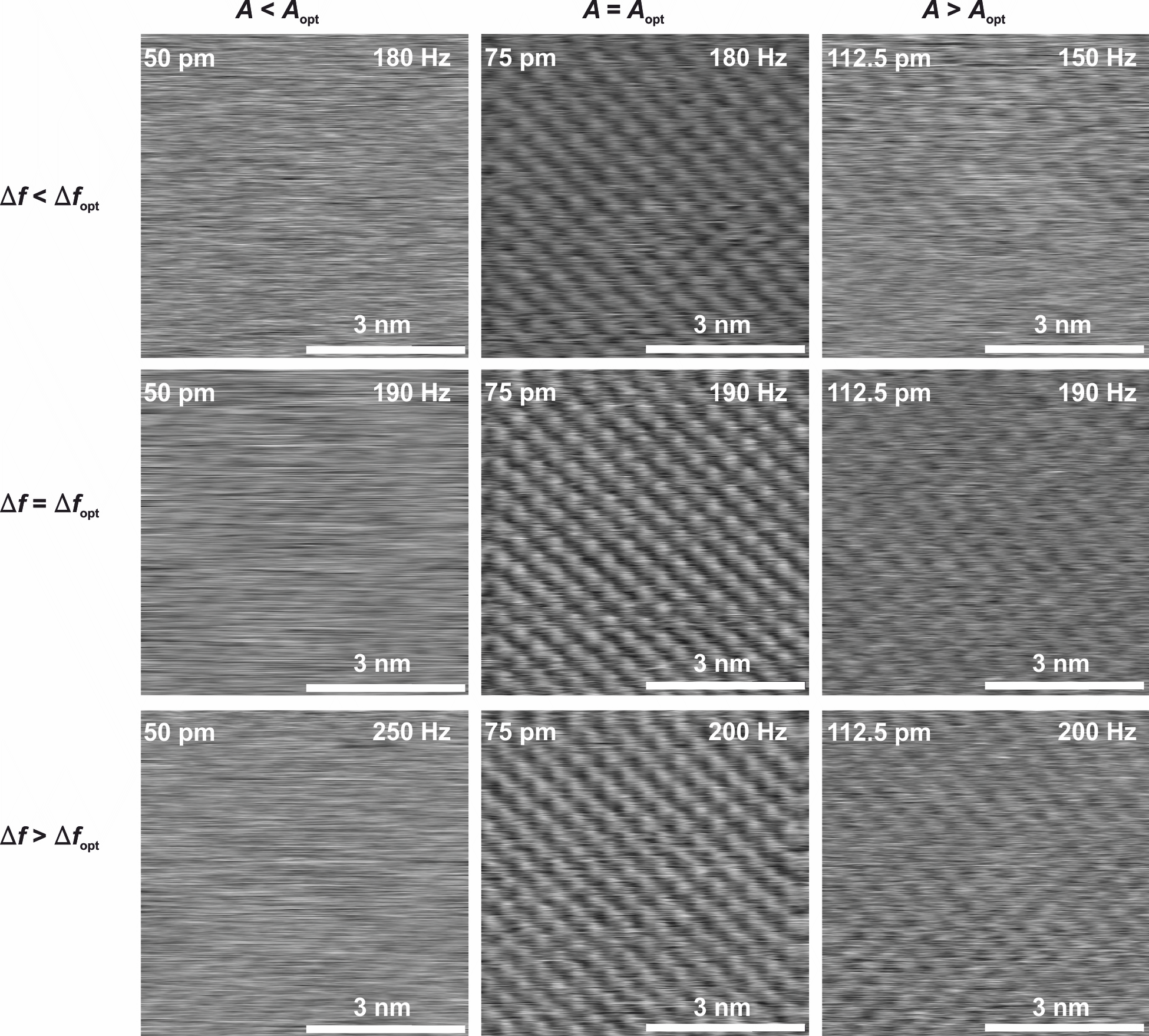}
\caption{\label{fAopt} Compilation of topographic images of the KBr(001) surface plane with amplitude and frequency shift set point shown in each image.  The center of the image shows  data with $A_{\mathrm{opt}}=75\,$pm and $\Delta f_{\mathrm{opt}}=190\,$Hz.
An increased frequency shift  $\Delta f > \Delta f_{\mathrm{opt}}$ leads to a higher rate of tip changes. Parameters of qPlus sensor ($k = 1000\,\frac{\text{N}}{\text{m}}$, $f_0=38853\,$Hz) equipped with a sapphire tip. Image processing: line flattening.}
\end{figure*}

Finally we discuss the effect of  frequency shift set point  $\Delta f$  on imaging in Fig.~\ref{fAopt}. The center image is the optimal case with $A_{\mathrm{opt}}=75\,$pm and a frequency shift of  $\Delta f_{\mathrm{opt}}=190\,$Hz. In the surrounding images, the amplitude and the frequency shift are varied.
All three amplitude set points: $A < A_{\mathrm{opt}}$, $A = A_{\mathrm{opt}}$ and $A > A_{\mathrm{opt}}$ share a decrease in the contrast due to a decreasing in signal when lowering the frequency shift ($\Delta f<\Delta f_{\mathrm{opt}}$). For higher frequency shifts $\Delta f >  \Delta f_{\mathrm{opt}}$, tip changes become more frequent. The data in Fig.~\ref{fAopt} demonstrates that the optimal frequency shift $\Delta f_{\mathrm{opt}}$ is a compromise between an ideal SNR and an acceptable rate of tip changes, and that the oscillation amplitude can be freely adjusted to optimize the SNR.

\section{Conclusion}

We have demonstrated  atomic resolution in ambient conditions and analyzed the main problems one has to deal with in uncontrolled environment. The water layer which forms on any surface was imaged and the height of a single water layer was measured to be 200\,pm, comparable to observations of other groups. High damping due to the hydration layers was demonstrated, and the effect on damping was shown by contrasting relatively wet and dry surfaces.
We further characterized the damping by recording drive signal versus amplitude spectra and deriving the effective damping $Q_{\mathrm{eff}}$.
This we used to systematically obtain the optimal imaging parameters for different sensors and and tip materials.

\section{Acknowledgments}
We thank Bill Muller and Veeco Metrology (now Bruker), Santa Barbara, USA for donating the microscope head we used in our study as well as Elisabeth Wutscher for preliminary studies on the subject.
Financial support from the Deutsche Forschungsgemeinschaft (GRK 1570, SFB 689) is gratefully acknowledged.


\begin{thebibliography}{99}
\expandafter\ifx\csname natexlab\endcsname\relax\def\natexlab#1{#1}\fi
\expandafter\ifx\csname bibnamefont\endcsname\relax
  \def\bibnamefont#1{#1}\fi
\expandafter\ifx\csname bibfnamefont\endcsname\relax
  \def\bibfnamefont#1{#1}\fi
\expandafter\ifx\csname citenamefont\endcsname\relax
  \def\citenamefont#1{#1}\fi
\expandafter\ifx\csname url\endcsname\relax
  \def\url#1{\texttt{#1}}\fi
\expandafter\ifx\csname urlprefix\endcsname\relax\def\urlprefix{URL }\fi
\providecommand{\bibinfo}[2]{#2}
\providecommand{\eprint}[2][]{\url{#2}}

\bibitem[{\citenamefont{Binnig and Quate}(1986)}]{Binnig1986a}
\bibinfo{author}{\bibfnamefont{G.}~\bibnamefont{Binnig}},
  \bibinfo{author}{\bibfnamefont{C.~F.} \bibnamefont{Quate}} \bibnamefont{and}
   \bibinfo{author}{\bibfnamefont{Ch.} \bibnamefont{Gerber}},
  \bibinfo{journal}{Phys. Rev. Lett.} \textbf{\bibinfo{volume}{56}},
  \bibinfo{pages}{930} (\bibinfo{year}{1986}).

\bibitem[{\citenamefont{Albrecht et~al.}(1991)\citenamefont{Albrecht,
  Grutter, Horne, and Rugar}}]{Albrecht1991a}
\bibinfo{author}{\bibfnamefont{T.~R.} \bibnamefont{Albrecht}},
  \bibinfo{author}{\bibfnamefont{P.}~\bibnamefont{Grutter}},
  \bibinfo{author}{\bibfnamefont{D.}~\bibnamefont{Horne}} \bibnamefont{and}
  \bibinfo{author}{\bibfnamefont{D.}~\bibnamefont{Rugar}}, \bibinfo{journal}{J.
  Appl. Phys.} \textbf{\bibinfo{volume}{69}}, \bibinfo{pages}{668}
  (\bibinfo{year}{1991}).


  \bibitem[{\citenamefont{Garcia}(2002)}]{Garcia2002}
  \bibinfo{author}{\bibfnamefont{R.} \bibnamefont{Garcia}},
  \bibinfo{journal}{Surf. Sci. Rep.} \textbf{\bibinfo{volume}{47}},
  \bibinfo{pages}{197} (\bibinfo{year}{2002}).



 \bibitem[{\citenamefont{Morita} \emph{et~al.}(2002)\citenamefont{Morita, Meyer and Wiesendanger}}]{Morita2002bk}
   \bibinfo{year}{2002}, in \emph{\bibinfo{booktitle}{Noncontact Atomic Force
  Microscopy}}, edited by
  \bibinfo{editor}{\bibfnamefont{S.}~\bibnamefont{Morita}},
  \bibinfo{editor}{\bibfnamefont{R.}~\bibnamefont{Wiesendanger}} \bibnamefont{and}
  \bibinfo{editor}{\bibfnamefont{E.}~\bibnamefont{Meyer}}
  (\bibinfo{publisher}{Springer Berlin Heidelberg New York}).


  \bibitem[{\citenamefont{Giessibl}(2003)}]{Giessibl2003}
  \bibinfo{author}{\bibfnamefont{F.~J.} \bibnamefont{Giessibl}},
  \bibinfo{journal}{Rev. Mod. Phys.} \textbf{\bibinfo{volume}{75}},
  \bibinfo{pages}{949} (\bibinfo{year}{2003}).

\bibitem[{\citenamefont{Morita} \emph{et~al.}(2009)\citenamefont{Morita, Meyer and Wiesendanger}}]{Morita2009bk}
   \bibinfo{year}{2002}, in \emph{\bibinfo{booktitle}{Noncontact Atomic Force
  Microscopy II}}, edited by
  \bibinfo{editor}{\bibfnamefont{S.}~\bibnamefont{Morita}},
  \bibinfo{editor}{\bibfnamefont{F.J.}~\bibnamefont{Giessibl}} \bibnamefont{and}
  \bibinfo{editor}{\bibfnamefont{R.}~\bibnamefont{Wiesendanger}}
  (\bibinfo{publisher}{Springer Berlin Heidelberg New York}).

\bibitem[{\citenamefont{Kodera et~al.}(2010)\citenamefont{Kodera, Yamamoto,
  Ishikawa, and Ando}}]{Kodera2010}
\bibinfo{author}{\bibfnamefont{N.}~\bibnamefont{Kodera}},
  \bibinfo{author}{\bibfnamefont{D.}~\bibnamefont{Yamamoto}},
  \bibinfo{author}{\bibfnamefont{R.}~\bibnamefont{Ishikawa}} \bibnamefont{and}
  \bibinfo{author}{\bibfnamefont{T.}~\bibnamefont{Ando}},
  \bibinfo{journal}{Nature} \textbf{\bibinfo{volume}{468}}, \bibinfo{pages}{72}
  (\bibinfo{year}{2010}).


\bibitem[{\citenamefont{Binnig and Rohrer}(1983)}]{BINNIG1983}
\bibinfo{author}{\bibfnamefont{G.}~\bibnamefont{Binnig}} \bibnamefont{and}
  \bibinfo{author}{\bibfnamefont{H.}~\bibnamefont{Rohrer}},
  \bibinfo{journal}{Surf. Sci.} \textbf{\bibinfo{volume}{126}},
  \bibinfo{pages}{236} (\bibinfo{year}{1983}).


  \bibitem[{\citenamefont{Bammerlin et~al.}(1998)\citenamefont{Bammerlin,
  L\"{u}thi, Meyer, Baratoff, L\"{u}, Guggisberg, Loppacher, Gerber, and
  G\"{u}ntherodt}}]{Bammerlin1998}
\bibinfo{author}{\bibfnamefont{M.}~\bibnamefont{Bammerlin}},
  \bibinfo{author}{\bibfnamefont{R.}~\bibnamefont{L\"{u}thi}},
  \bibinfo{author}{\bibfnamefont{E.}~\bibnamefont{Meyer}},
  \bibinfo{author}{\bibfnamefont{A.}~\bibnamefont{Baratoff}},
  \bibinfo{author}{\bibfnamefont{J.}~\bibnamefont{L\"{u}}},
  \bibinfo{author}{\bibfnamefont{M.}~\bibnamefont{Guggisberg}},
  \bibinfo{author}{\bibfnamefont{C.}~\bibnamefont{Loppacher}},
  \bibinfo{author}{\bibfnamefont{C.}~\bibnamefont{Gerber}} \bibnamefont{and}
  \bibinfo{author}{\bibfnamefont{H.-J.} \bibnamefont{G\"{u}ntherodt}},
  \bibinfo{journal}{Appl. Phys. A} \textbf{\bibinfo{volume}{66}},
  \bibinfo{pages}{S293} (\bibinfo{year}{1998}).




\bibitem[{\citenamefont{Rasmussen et~al.}(1986)}]{Rasmussen1998a}
\bibinfo{author}{\bibfnamefont{P. B.}~\bibnamefont{Rasmussen}} \bibnamefont,
  \bibinfo{author}{\bibfnamefont{B. L. M.}~\bibnamefont{Hendriksen}},
   \bibinfo{author}{\bibfnamefont{H.}~\bibnamefont{Zeijlemaker}},
    \bibinfo{author}{\bibfnamefont{H. G.}~\bibnamefont{Ficke}} \bibnamefont{and}
     \bibinfo{author}{\bibfnamefont{J. W. M.} \bibnamefont{Frenken}},
  \bibinfo{journal}{Rev. Sci. Instrum.} \textbf{\bibinfo{volume}{69}},
  \bibinfo{pages}{3879} (\bibinfo{year}{1998}).




\bibitem[{\citenamefont{Wutscher and Giessibl}(2011)}]{Wutscher2011}
\bibinfo{author}{\bibfnamefont{E.}~\bibnamefont{Wutscher}} \bibnamefont{and}
  \bibinfo{author}{\bibfnamefont{F.~J.} \bibnamefont{Giessibl}},
  \bibinfo{journal}{Rev. Sci. Inst.} \textbf{\bibinfo{volume}{82}},
  \bibinfo{pages}{093703} (\bibinfo{year}{2011}).



 \bibitem[{\citenamefont{Labuda et~al.}(2013)\citenamefont{Labuda, Kobayashi, Suzuki, Yamada and Grütter}}]{Labuda2013}
\bibinfo{author}{\bibfnamefont{A.}~\bibnamefont{Labuda}},
  \bibinfo{author}{\bibfnamefont{K.}~\bibnamefont{Kobayashi}},
  \bibinfo{author}{\bibfnamefont{K.}~\bibnamefont{Suzuki}},
  \bibinfo{author}{\bibfnamefont{H.}~\bibnamefont{Yamada}} \bibnamefont{and}
  \bibinfo{author}{\bibfnamefont{P.} \bibnamefont{Grutter}},
  \bibinfo{journal}{Physi. Rev. Lett.} \textbf{\bibinfo{volume}{110}},
  \bibinfo{pages}{066102} (\bibinfo{year}{2013}).


  \bibitem[{\citenamefont{Israelachvili}(1991)}]{Israelachvili1991}
\bibinfo{author}{\bibnamefont{Israelachvili}, \bibfnamefont{J.}},
  \bibinfo{year}{1991}, \emph{\bibinfo{title}{Intermolecular and Surface
  Forces, 2nd ed.}} (\bibinfo{publisher}{{}Academic Press, London}).





\bibitem[{\citenamefont{Ohnesorge}(2005)\citenamefont{Fukuma, Kobayashi,
  Matsushige, and Yamada}}]{Ohnesorge1993}
\bibinfo{author}{\bibfnamefont{F.}~\bibnamefont{Ohnesorge}} \bibnamefont{and}
  \bibinfo{author}{\bibfnamefont{G.}~\bibnamefont{Binnig}},
  \bibinfo{journal}{Science} \textbf{\bibinfo{volume}{260}},
  \bibinfo{pages}{1451} (\bibinfo{year}{1993}).

\bibitem[{\citenamefont{Fukuma et~al.}(2005)\citenamefont{Fukuma, Kobayashi,
  Matsushige, and Yamada}}]{Fukuma2005b}
\bibinfo{author}{\bibfnamefont{T.}~\bibnamefont{Fukuma}},
  \bibinfo{author}{\bibfnamefont{K.}~\bibnamefont{Kobayashi}},
  \bibinfo{author}{\bibfnamefont{K.}~\bibnamefont{Matsushige}} \bibnamefont{and}
  \bibnamefont{and} \bibinfo{author}{\bibfnamefont{H.}~\bibnamefont{Yamada}},
  \bibinfo{journal}{Appl. Phys. Lett.} \textbf{\bibinfo{volume}{87}},
  \bibinfo{pages}{034101} (\bibinfo{year}{2005}).

\bibitem[{\citenamefont{Rhode et~al.}(2009)\citenamefont{Rode, Oyabu, Kobayashi,
  Yamada, and K\"{u}hnle}}]{Rhode2009a}
\bibinfo{author}{\bibfnamefont{S.}~\bibnamefont{Rhode}},
  \bibinfo{author}{\bibfnamefont{N.}~\bibnamefont{Oyabu}},
  \bibinfo{author}{\bibfnamefont{K.}~\bibnamefont{Kobayashi}},
  \bibinfo{author}{\bibfnamefont{H.}~\bibnamefont{Yamada}} \bibnamefont{and}
  \bibinfo{author}{\bibfnamefont{A.}~\bibnamefont{K\"{u}hnle}},
  \bibinfo{journal}{Langmuir} \textbf{\bibinfo{volume}{25}},
  \bibinfo{pages}{2850} (\bibinfo{year}{2009}).

\bibitem[{\citenamefont{Hoogenboom et~al.}(2006)\citenamefont{Hoogenboom, Hug,
  Pellmont, Martin, Frederix, Fotiadis, and Engel}}]{Hoogenboom2006}
\bibinfo{author}{\bibfnamefont{B.~W.} \bibnamefont{Hoogenboom}},
  \bibinfo{author}{\bibfnamefont{H.~J.} \bibnamefont{Hug}},
  \bibinfo{author}{\bibfnamefont{Y.}~\bibnamefont{Pellmont}},
  \bibinfo{author}{\bibfnamefont{S.}~\bibnamefont{Martin}},
  \bibinfo{author}{\bibfnamefont{P.~L. T.~M.} \bibnamefont{Frederix}},
  \bibinfo{author}{\bibfnamefont{D.}~\bibnamefont{Fotiadis}} \bibnamefont{and}
  \bibinfo{author}{\bibfnamefont{A.}~\bibnamefont{Engel}},
  \bibinfo{journal}{Appl. Phys. Lett.} \textbf{\bibinfo{volume}{88}},
  \bibinfo{pages}{193109} (\bibinfo{year}{2006}).

\bibitem[{\citenamefont{Ichii et~al.}(2012)\citenamefont{Ichii, Fujimura, Negami, Murase, Sugimura}}]{Ichii2012}
\bibinfo{author}{\bibfnamefont{T.}~\bibnamefont{Ichii}},
  \bibinfo{author}{\bibfnamefont{M.}~\bibnamefont{Fujimura}},
  \bibinfo{author}{\bibfnamefont{M.}~\bibnamefont{Negami}},
  \bibinfo{author}{\bibfnamefont{K.}~\bibnamefont{Murase}} \bibnamefont{and}
  \bibinfo{author}{\bibfnamefont{H.} \bibnamefont{Sugimura}},
  \bibinfo{journal}{Jap. J. Appl. Phys.} \textbf{\bibinfo{volume}{51}},
  \bibinfo{pages}{08KB08} (\bibinfo{year}{2012}).

\bibitem{Welker}
\bibinfo{author}{\bibfnamefont{J.} \bibnamefont{Welker}},
  \bibinfo{author}{\bibfnamefont{F.}~\bibnamefont{de Faria Elsner}} \bibnamefont{and}
  \bibinfo{author}{\bibfnamefont{F. J.}~\bibnamefont{Giessibl}}, \bibinfo{journal}{Appl. Phys. Lett.} \textbf{\bibinfo{volume}{99}}, \bibinfo{pages}{084102}
  (\bibinfo{year}{2011}).


\bibitem[{\citenamefont{Giessibl}(2000)}]{Giessibl2000}
\bibinfo{author}{\bibfnamefont{F.~J.} \bibnamefont{Giessibl}},
  \bibinfo{journal}{Appl. Phys. Lett.} \textbf{\bibinfo{volume}{76}},
  \bibinfo{pages}{1470} (\bibinfo{year}{2000}).



 \bibitem[{\citenamefont{Giessibl et~al.}(2011)\citenamefont{Giessibl,
  Pielmeier, Eguchi, An, and Hasegawa}}]{Giessibl2011a}
\bibinfo{author}{\bibfnamefont{F.~J.}~\bibnamefont{Giessibl}},
  \bibinfo{author}{\bibfnamefont{F.}~\bibnamefont{Pielmeier}},
  \bibinfo{author}{\bibfnamefont{T.}~\bibnamefont{Eguchi}},
  \bibinfo{author}{\bibfnamefont{T.}~\bibnamefont{An}} \bibnamefont{and}
  \bibinfo{author}{\bibfnamefont{Y.}~\bibnamefont{Hasegawa}},
  \bibinfo{journal}{Phys. Rev. B} \textbf{\bibinfo{volume}{84}},
  \bibinfo{pages}{125409} (\bibinfo{year}{2011}).


\bibitem[{\citenamefont{Anczykowski}
  \emph{et~al.}(1999)\citenamefont{Anczykowski, Gotsmann, Fuchs, Cleveland, and
  Elings}}]{Anczykowski1999}
\bibinfo{author}{\bibnamefont{Anczykowski}, \bibfnamefont{B.}},
  \bibinfo{author}{\bibfnamefont{B.}~\bibnamefont{Gotsmann}},
  \bibinfo{author}{\bibfnamefont{H.}~\bibnamefont{Fuchs}},
  \bibinfo{author}{\bibfnamefont{J.~P.} \bibnamefont{Cleveland}}  \bibnamefont{and}
  \bibinfo{author}{\bibfnamefont{V.~B.} \bibnamefont{Elings}},
  \bibinfo{journal}{{}Appl. Surf.
  Sci.} \textbf{\bibinfo{volume}{140}},  \bibinfo{pages}{376} (\bibinfo{year}{1999}).


\bibitem[{\citenamefont{Giessibl}(2004)}]{Giessibl2004}
\bibinfo{author}{\bibfnamefont{F.~J.} \bibnamefont{Giessibl}},
\bibinfo{author}{\bibfnamefont{S.} \bibnamefont{Hembacher}},
\bibinfo{author}{\bibfnamefont{M.} \bibnamefont{Herz}},
\bibinfo{author}{\bibfnamefont{Ch.} \bibnamefont{Schiller}},
\bibinfo{author}{\bibfnamefont{J.} \bibnamefont{Mannhart}},
  \bibinfo{journal}{Nanotechnology} \textbf{\bibinfo{volume}{15}},
  \bibinfo{pages}{S79} (\bibinfo{year}{2004}).



\bibitem[{\citenamefont{Giessibl}(1998)}]{Giessibl1998}
\bibinfo{author}{\bibfnamefont{F.~J.} \bibnamefont{Giessibl}},
  \bibinfo{journal}{Appl. Phys. Lett.} \textbf{\bibinfo{volume}{73}},
  \bibinfo{pages}{3956} (\bibinfo{year}{1998}).






\bibitem[{\citenamefont{Giessibl and Trafas}(1994)}]{Giessibl1994a}
\bibinfo{author}{\bibnamefont{Giessibl}, \bibfnamefont{F.~J.}} and
  \bibinfo{author}{\bibfnamefont{B.~M.} \bibnamefont{Trafas}},
     \bibinfo{journal}{Rev. Sci. Instrum.} \textbf{\bibinfo{volume}{65}},
  \bibinfo{pages}{1923} (\bibinfo{year}{1994}).





\bibitem[{\citenamefont{Meyer and Amer}(1990)}]{Meyer1990}
\bibinfo{author}{\bibfnamefont{G.}~\bibnamefont{Meyer}} \bibnamefont{and}
  \bibinfo{author}{\bibfnamefont{N.~M.} \bibnamefont{Amer}},
  \bibinfo{journal}{Appl. Phys. Lett.} \textbf{\bibinfo{volume}{56}},
  \bibinfo{pages}{2100} (\bibinfo{year}{1990}).



\bibitem[{\citenamefont{Meyer et~al.}(1990)\citenamefont{Meyer, Heinzelmann,
  Rudin, and G\"{u}ntherodt}}]{Meyer1990a}
\bibinfo{author}{\bibfnamefont{E.}~\bibnamefont{Meyer}},
  \bibinfo{author}{\bibfnamefont{H.}~\bibnamefont{Heinzelmann}},
  \bibinfo{author}{\bibfnamefont{H.}~\bibnamefont{Rudin}} \bibnamefont{and}
  \bibinfo{author}{\bibfnamefont{H.~J.} \bibnamefont{G\"{u}ntherodt}},
  \bibinfo{journal}{Z. Phys. B} \textbf{\bibinfo{volume}{79}},
  \bibinfo{pages}{3} (\bibinfo{year}{1990}).

 \bibitem[{\citenamefont{Giessibl and Binnig}(1992)}]{Giessibl1992b}
\bibinfo{author}{\bibfnamefont{F.~J.}~\bibnamefont{Giessibl}} \bibnamefont{and}
  \bibinfo{author}{\bibfnamefont{G.}~\bibnamefont{Binnig}},
  \bibinfo{journal}{Ultramicroscopy} \textbf{\bibinfo{volume}{42-44}},
  \bibinfo{pages}{281} (\bibinfo{year}{1992}).





\bibitem[{\citenamefont{Bennewitz et~al.}(2002)\citenamefont{Bennewitz,
  Pfeiffer, Sch\"{a}r, Barwich, Meyer, and Kantorovich}}]{Bennewitz2002}
\bibinfo{author}{\bibfnamefont{R.}~\bibnamefont{Bennewitz}},
  \bibinfo{author}{\bibfnamefont{O.}~\bibnamefont{Pfeiffer}},
  \bibinfo{author}{\bibfnamefont{S.}~\bibnamefont{Sch\"{a}r}},
  \bibinfo{author}{\bibfnamefont{V.}~\bibnamefont{Barwich}},
  \bibinfo{author}{\bibfnamefont{E.}~\bibnamefont{Meyer}} \bibnamefont{and}
  \bibinfo{author}{\bibfnamefont{L.}~\bibnamefont{Kantorovich}},
  \bibinfo{journal}{Appl. Surf. Sci.} \textbf{\bibinfo{volume}{188}},
  \bibinfo{pages}{232} (\bibinfo{year}{2002}).

\bibitem[{\citenamefont{Maier et~al.}(2007)\citenamefont{Maier, Pfeiffer,
  Glatzel, Meyer, Filleter, and Bennewitz}}]{Maier2007}
\bibinfo{author}{\bibfnamefont{S.}~\bibnamefont{Maier}},
  \bibinfo{author}{\bibfnamefont{O.}~\bibnamefont{Pfeiffer}},
  \bibinfo{author}{\bibfnamefont{T.}~\bibnamefont{Glatzel}},
  \bibinfo{author}{\bibfnamefont{E.}~\bibnamefont{Meyer}},
  \bibinfo{author}{\bibfnamefont{T.}~\bibnamefont{Filleter}} \bibnamefont{and}
  \bibinfo{author}{\bibfnamefont{R.}~\bibnamefont{Bennewitz}},
  \bibinfo{journal}{Phys. Rev. B} \textbf{\bibinfo{volume}{75}},
  \bibinfo{pages}{195408} (\bibinfo{year}{2007}).



  \bibitem[{\citenamefont{Pawlak et~al.}(2012)\citenamefont{Pawlak, Kawai, Fremy,
  Glatzel, and Meyer}}]{Pawlak2012}
\bibinfo{author}{\bibfnamefont{R.}~\bibnamefont{Pawlak}},
  \bibinfo{author}{\bibfnamefont{S.}~\bibnamefont{Kawai}},
  \bibinfo{author}{\bibfnamefont{S.}~\bibnamefont{Fremy}},
  \bibinfo{author}{\bibfnamefont{T.}~\bibnamefont{Glatzel}} \bibnamefont{and}
  \bibinfo{author}{\bibfnamefont{E.}~\bibnamefont{Meyer}},
  \bibinfo{journal}{JPCM} \textbf{\bibinfo{volume}{24}},
  \bibinfo{pages}{084005} (\bibinfo{year}{2012}).


  \bibitem[{\citenamefont{Repp et~al.}(2004)\citenamefont{Repp, Meyer, Olsson, Persson}}]{Repp2004}
\bibinfo{author}{\bibfnamefont{J.}~\bibnamefont{Repp}},
\bibinfo{author}{\bibfnamefont{G.}~\bibnamefont{Meyer}},
\bibinfo{author}{\bibfnamefont{F. E.}~\bibnamefont{Olsson}} \bibnamefont{and}
  \bibinfo{author}{\bibfnamefont{M.}~\bibnamefont{Persson}},
  \bibinfo{journal}{Sience} \textbf{\bibinfo{volume}{305}},
  \bibinfo{pages}{493} (\bibinfo{year}{2004}).

 \bibitem[{\citenamefont{Gross et~al.}(2009)\citenamefont{Gross, Mohn, Moll, Liljeroth and Meyer}}]{Gross2009}
\bibinfo{author}{\bibfnamefont{L.}~\bibnamefont{Gross}},
\bibinfo{author}{\bibfnamefont{F.}~\bibnamefont{Mohn}},
\bibinfo{author}{\bibfnamefont{N.}~\bibnamefont{Moll}},
  \bibinfo{author}{\bibfnamefont{P.}~\bibnamefont{Liljeroth}} \bibnamefont{and}
  \bibinfo{author}{\bibfnamefont{G.}~\bibnamefont{Meyer}},
  \bibinfo{journal}{Sience} \textbf{\bibinfo{volume}{325}},
  \bibinfo{pages}{1110} (\bibinfo{year}{2009}).



   \bibitem[{\citenamefont{Repp et~al.}(2005)\citenamefont{Repp and Meyer}}]{Repp2005}
\bibinfo{author}{\bibfnamefont{J.}~\bibnamefont{Repp}},
  \bibinfo{author}{\bibfnamefont{G.}~\bibnamefont{Meyer}},
 \bibinfo{author}{\bibfnamefont{S. M.}~\bibnamefont{Stojkovic}},
  \bibinfo{author}{\bibfnamefont{A.}~\bibnamefont{Gourdon}} \bibnamefont{and}
  \bibinfo{author}{\bibfnamefont{C.}~\bibnamefont{Joachim}},
  \bibinfo{journal}{Phys. Rev. Lett.} \textbf{\bibinfo{volume}{94}},
  \bibinfo{pages}{026803} (\bibinfo{year}{2005}).



   \bibitem[{\citenamefont{Pavlicek et~al.}(2012)\citenamefont{Repp and Meyer}}]{Pavlicek2012}
\bibinfo{author}{\bibfnamefont{N.}~\bibnamefont{Pavlicek}},
\bibinfo{author}{\bibfnamefont{B.}~\bibnamefont{Fleury}},
\bibinfo{author}{\bibfnamefont{M.}~\bibnamefont{Neu}},
\bibinfo{author}{\bibfnamefont{J.}~\bibnamefont{Niedenf{\"u}hr}},
 \bibinfo{author}{\bibfnamefont{C.}~\bibnamefont{Herranz-Lancho}},
 \bibinfo{author}{\bibfnamefont{M.}~\bibnamefont{Ruben}} \bibnamefont{and}
 \bibinfo{author}{\bibfnamefont{J.}~\bibnamefont{Repp}},
  \bibinfo{journal}{Phys. Rev. Lett.} \textbf{\bibinfo{volume}{108}},
  \bibinfo{pages}{086101 } (\bibinfo{year}{2012}).


 \bibitem[{\citenamefont{Gross et~al.}(2010)\citenamefont{Gross, Mohn, Moll, Meyer, Ebel, Abdel-Mageed and Jaspars }}]{Gross2010}
\bibinfo{author}{\bibfnamefont{L.}~\bibnamefont{Gross}},
\bibinfo{author}{\bibfnamefont{F.}~\bibnamefont{Mohn}},
\bibinfo{author}{\bibfnamefont{N.}~\bibnamefont{Moll}},
  \bibinfo{author}{\bibfnamefont{G.}~\bibnamefont{Meyer}},
    \bibinfo{author}{\bibfnamefont{R.}~\bibnamefont{Ebel}},
  \bibinfo{author}{\bibfnamefont{W. M.}~\bibnamefont{Abdel-Mageed}} \bibnamefont{and}
  \bibinfo{author}{\bibfnamefont{M.}~\bibnamefont{Jaspars}},
  \bibinfo{journal}{Nature chemistry} \textbf{\bibinfo{volume}{10}},
  \bibinfo{pages}{821} (\bibinfo{year}{2010})





   \bibitem[{\citenamefont{Filleter et~al.}(2012)\citenamefont{Filleter, Paul and Bennewitz}}]{Filleter2008}
\bibinfo{author}{\bibfnamefont{T.}~\bibnamefont{Filleter}},
\bibinfo{author}{\bibfnamefont{W.}~\bibnamefont{Paul}} \bibnamefont{and}
  \bibinfo{author}{\bibfnamefont{R.}~\bibnamefont{Bennewitz}},
  \bibinfo{journal}{Phys. Rev. B.} \textbf{\bibinfo{volume}{77}},
  \bibinfo{pages}{035430} (\bibinfo{year}{2008}).












  \bibitem[{\citenamefont{Hofer et~al.}(2003)\citenamefont{Hofer, Foster, and
  Shluger}}]{Hofer2003}
\bibinfo{author}{\bibfnamefont{W.}~\bibnamefont{Hofer}},
  \bibinfo{author}{\bibfnamefont{A.}~\bibnamefont{Foster}} \bibnamefont{and}
  \bibinfo{author}{\bibfnamefont{A.}~\bibnamefont{Shluger}},
  \bibinfo{journal}{Rev. Mod. Phys.} \textbf{\bibinfo{volume}{75}},
  \bibinfo{pages}{1287} (\bibinfo{year}{2003}).





\bibitem{Palasantzas2009}
\bibinfo{author}{\bibfnamefont{G.} \bibnamefont{Palasantzas}},
  \bibinfo{author}{\bibfnamefont{V. B.}~\bibnamefont{Svetovoy}} \bibnamefont{and}
  \bibinfo{author}{\bibfnamefont{P. J.}~\bibnamefont{van Zwol}},
 \bibinfo{journal}{Phys. Rev. B} \textbf{\bibinfo{volume}{79}} (23), \bibinfo{pages}{235434}
  (\bibinfo{year}{2009}).



\bibitem{James2011}
\bibinfo{author}{\bibfnamefont{M.} \bibnamefont{James}},
  \bibinfo{author}{\bibfnamefont{T. A.}~\bibnamefont{Darwish}},
  \bibinfo{author}{\bibfnamefont{S.}~\bibnamefont{Ciampi}},
  \bibinfo{author}{\bibfnamefont{S. O.}~\bibnamefont{Sylvester}},
  \bibinfo{author}{\bibfnamefont{Z.}~\bibnamefont{Zhang}},
  \bibinfo{author}{\bibfnamefont{A.}~\bibnamefont{Ng}},
   \bibinfo{author}{\bibfnamefont{J. J.}~\bibnamefont{Gooding}} \bibnamefont{and}
  \bibinfo{author}{\bibfnamefont{T. L.}~\bibnamefont{Hanley}},
 \bibinfo{journal}{Soft Matter} \textbf{\bibinfo{volume}{7}} , \bibinfo{pages}{5309}
  (\bibinfo{year}{2011}).

\bibitem{Davy1998}
\bibinfo{author}{\bibfnamefont{S.} \bibnamefont{Davy}},
  \bibinfo{author}\bibfnamefont{M.}~\bibnamefont{Spajer} \bibnamefont{and}
 \bibinfo{author}{\bibfnamefont{D.}~\bibnamefont{Courjon}},
 \bibinfo{journal}{Appl. Phys. Lett.} \textbf{\bibinfo{volume}{73}} , \bibinfo{pages}{2594}
  (\bibinfo{year}{1998}).

  \bibitem{Wei2000}
\bibinfo{author}{\bibfnamefont{P. K.} \bibnamefont{Wei}} \bibnamefont{and}
 \bibinfo{author}{\bibfnamefont{W. S.}~\bibnamefont{Fann}},
 \bibinfo{journal}{J. Appl. Phys.} \textbf{\bibinfo{volume}{87}} , \bibinfo{pages}{2561}
  (\bibinfo{year}{2000}).

\bibitem{Huang2007}
\bibinfo{author}{\bibfnamefont{F. M.} \bibnamefont{Huang}},
  \bibinfo{author}\bibfnamefont{F.}~\bibnamefont{Culfaz}  \bibinfo{author}{\bibfnamefont{F.}~\bibnamefont{Festy}} \bibnamefont{and}
 \bibinfo{author}{\bibfnamefont{D.}~\bibnamefont{Richards}},
 \bibinfo{journal}{Nanotechnology} \textbf{\bibinfo{volume}{18}}, \bibinfo{pages}{015501}
  (\bibinfo{year}{2007}).

\bibitem{Freund1999}
\bibinfo{author}{\bibfnamefont{J.} \bibnamefont{Freund}},
  \bibinfo{author}\bibfnamefont{J.}~\bibnamefont{Halbritter} \bibnamefont{and}
 \bibinfo{author}{\bibfnamefont{J. K. H.}~\bibnamefont{H{\"o}rber}},
 \bibinfo{journal}{Micro. Res. Tech.} \textbf{\bibinfo{volume}{44}} , \bibinfo{pages}{327}
  (\bibinfo{year}{1999}).




\bibitem[{\citenamefont{Filleter et~al.}(2006)\citenamefont{Filleter, Maier,
  and Bennewitz}}]{Filleter2006}
\bibinfo{author}{\bibfnamefont{T.}~\bibnamefont{Filleter}},
  \bibinfo{author}{\bibfnamefont{S.}~\bibnamefont{Maier}} \bibnamefont{and}
  \bibinfo{author}{\bibfnamefont{R.}~\bibnamefont{Bennewitz}},
  \bibinfo{journal}{Phys. Rev. B} \textbf{\bibinfo{volume}{73}},
  \bibinfo{pages}{155433} (\bibinfo{year}{2006}).



  \bibitem[{\citenamefont{Luna et~al.}(1998)\citenamefont{Luna, Rieutord, Melman,
  Dai, and Salmeron}}]{Luna1998}
\bibinfo{author}{\bibfnamefont{M.}~\bibnamefont{Luna}},
  \bibinfo{author}{\bibfnamefont{F.}~\bibnamefont{Rieutord}},
  \bibinfo{author}{\bibfnamefont{N.~A.} \bibnamefont{Melman}},
  \bibinfo{author}{\bibfnamefont{Q.}~\bibnamefont{Dai}} \bibnamefont{and}
  \bibinfo{author}{\bibfnamefont{M.}~\bibnamefont{Salmeron}},
  \bibinfo{journal}{J. Phys. Chem. A} \textbf{\bibinfo{volume}{102}},
  \bibinfo{pages}{6793} (\bibinfo{year}{1998}).


  \bibitem[{\citenamefont{Jeffery et~al.}(2004)\citenamefont{Jeffery, Hoffmann, Pethica, Ramanujan, OzgurOzer and Oral}}]{Jeffery2004}
\bibinfo{author}{\bibfnamefont{S.}~\bibnamefont{Jeffery}},
  \bibinfo{author}{\bibfnamefont{P.~M.}~\bibnamefont{Hoffmann}},
  \bibinfo{author}{\bibfnamefont{J.~B.}~\bibnamefont{Pethica}},
  \bibinfo{author}{\bibfnamefont{C.}~\bibnamefont{Ramanujan}},
   \bibinfo{author}{\bibfnamefont{H.}~\bibnamefont{Ozgur Ozer}} \bibnamefont{and}
  \bibinfo{author}{\bibfnamefont{A.}~\bibnamefont{Oral}},
  \bibinfo{journal}{Phys. Rev. B} \textbf{\bibinfo{volume}{70}},
  \bibinfo{pages}{054114} (\bibinfo{year}{2004}).

\bibitem[{\citenamefont{Fukuma et~al.}(2007)\citenamefont{Fukuma, Higgins and Jarvis}}]{Fukuma2007}
\bibinfo{author}{\bibfnamefont{T.}~\bibnamefont{Fukuma}},
  \bibinfo{author}{\bibfnamefont{M.~J.}~\bibnamefont{Higgins}} \bibnamefont{and}
  \bibinfo{author}{\bibfnamefont{S.~P.} \bibnamefont{Jarvis}},
  \bibinfo{journal}{Biophys. J.} \textbf{\bibinfo{volume}{92}},
  \bibinfo{pages}{3603} (\bibinfo{year}{2007}).


  \bibitem[{\citenamefont{Kimura et~al.}(2010)\citenamefont{Kimura, Ido, Oyabu, Kobayashi and Hirata}}]{Kimura2010}
\bibinfo{author}{\bibfnamefont{K.}~\bibnamefont{Kimura}},
  \bibinfo{author}{\bibfnamefont{S.}~\bibnamefont{Ido}},
  \bibinfo{author}{\bibfnamefont{N.}~\bibnamefont{Oyabu}},
  \bibinfo{author}{\bibfnamefont{K.}~\bibnamefont{Kobayashi}} \bibnamefont{and}
  \bibinfo{author}{\bibfnamefont{Y.} \bibnamefont{Hirata}},
  \bibinfo{journal}{J. Chem. Phys.} \textbf{\bibinfo{volume}{132}},
  \bibinfo{pages}{194705} (\bibinfo{year}{2010}).


 \bibitem[{\citenamefont{Israelachvili  et~al.}(1983)\citenamefont{Israelachvili and Pashley}}]{Israelachvili1983}
\bibinfo{author}{\bibfnamefont{J. N.}~\bibnamefont{Israelachvili}} \bibnamefont{and}
  \bibinfo{author}{\bibfnamefont{R. M.} \bibnamefont{ Pashley}},
  \bibinfo{journal}{Nature } \textbf{\bibinfo{volume}{306}},
  \bibinfo{pages}{249} (\bibinfo{year}{1983}).

\bibitem[{\citenamefont{Kilpatrick et~al.}(2013)\citenamefont{Kilpatrick, }}]{Kilpatrick2013}
\bibinfo{author}{\bibfnamefont{J. I.}~\bibnamefont{Kilpatrick}},
  \bibinfo{author}{\bibfnamefont{S.}~\bibnamefont{Loh}} \bibnamefont{and}
  \bibinfo{author}{\bibfnamefont{S. P.} \bibnamefont{Jarvis}},
  \bibinfo{journal}{J. Am. Chem. Soc.} \textbf{\bibinfo{volume}{135}},
  \bibinfo{pages}{2628} (\bibinfo{year}{2013}).




  \bibitem[{\citenamefont{Schneiderbauer
  et~al.}(2012)\citenamefont{Schneiderbauer, Wastl, and
  Giessibl}}]{Schneiderbauer2012}
\bibinfo{author}{\bibfnamefont{M.}~\bibnamefont{Schneiderbauer}},
  \bibinfo{author}{\bibfnamefont{D.}~\bibnamefont{Wastl}}, \bibnamefont{and}
  \bibinfo{author}{\bibfnamefont{F.~J.} \bibnamefont{Giessibl}},
  \bibinfo{journal}{Beilstein J. Nanotechnol.} \textbf{\bibinfo{volume}{3}},
  \bibinfo{pages}{174} (\bibinfo{year}{2012}).

  \bibitem[{\citenamefont{Miranda}(1998)\citenamefont{Miranda, Xu, Shen and
  Salmeron}}]{Miranda1998}
\bibinfo{author}{\bibfnamefont{P.~B.}~\bibnamefont{Miranda}},
\bibinfo{author}{\bibfnamefont{Lei}~\bibnamefont{Xu}},
\bibinfo{author}{\bibfnamefont{Y.~R.}~\bibnamefont{Shen}} \bibnamefont{and}
\bibinfo{author}{\bibfnamefont{M.} \bibnamefont{Salmeron}},
\bibinfo{journal}{Phys. Rev. Lett.} \textbf{\bibinfo{volume}{81}},
\bibinfo{pages}{5876} (\bibinfo{year}{1998}).


\bibitem[{\citenamefont{F{\"o}lsch}(1992)\citenamefont{F{\"o}lsch, Stock, Henzler}}]{Foelsch1992}
\bibinfo{author}{\bibfnamefont{S.}~\bibnamefont{F{\"o}lsch}},
\bibinfo{author}{\bibfnamefont{A.}~\bibnamefont{Stock}} \bibnamefont{and}
\bibinfo{author}{\bibfnamefont{M.}~\bibnamefont{Henzler}}
\bibinfo{journal}{Surf. Sci.} \textbf{\bibinfo{volume}{264}},
\bibinfo{pages}{65} (\bibinfo{year}{1992}).


\bibitem[{\citenamefont{Peters}(1997)\citenamefont{Peters and Ewing}}]{Peters1997}
\bibinfo{author}{\bibfnamefont{S. J.}~\bibnamefont{Peters}} \bibnamefont{and}
\bibinfo{author}{\bibfnamefont{G. E.} \bibnamefont{Ewing}}
\bibinfo{journal}{J. Phys. Chem. B} \textbf{\bibinfo{volume}{101}},
\bibinfo{pages}{10880} (\bibinfo{year}{1997}).


  \bibitem[{\citenamefont{Wassermann}(1993)\citenamefont{Wassermann, Mirbt, Reif, Zink and
  Matthias}}]{Wassermann1993}
\bibinfo{author}{\bibfnamefont{B.}~\bibnamefont{Wassermann}},
\bibinfo{author}{\bibfnamefont{S.}~\bibnamefont{Mirbt}},
\bibinfo{author}{\bibfnamefont{J.}~\bibnamefont{Reif}}
\bibinfo{author}{\bibfnamefont{J. C.} \bibnamefont{Zink}} \bibnamefont{and}
\bibinfo{author}{\bibfnamefont{E. J.} \bibnamefont{Matthias}}
\bibinfo{journal}{J. Chem. Phys.} \textbf{\bibinfo{volume}{98}},
\bibinfo{pages}{10049} (\bibinfo{year}{1993}).


\bibitem[{\citenamefont{Meyer}(2001)\citenamefont{Meyer, Entel and  Hafner}}]{Meyer2001}
\bibinfo{author}{\bibfnamefont{H.}~\bibnamefont{Meyer}},
\bibinfo{author}{\bibfnamefont{P.}~\bibnamefont{Entel}} \bibnamefont{and}
\bibinfo{author}{\bibfnamefont{J.} \bibnamefont{Hafner}}
\bibinfo{journal}{Surf. Sci.} \textbf{\bibinfo{volume}{488}},
\bibinfo{pages}{177} (\bibinfo{year}{2001}).



 \bibitem[{\citenamefont{Park}(2004)\citenamefont{Park, Cho and  Kim}}]{Park2004}
\bibinfo{author}{\bibfnamefont{J. M.}~\bibnamefont{Park}},
\bibinfo{author}{\bibfnamefont{J. H.}~\bibnamefont{Cho}} \bibnamefont{and}
\bibinfo{author}{\bibfnamefont{K. S.} \bibnamefont{Kim}}
\bibinfo{journal}{Phys. Rev. B} \textbf{\bibinfo{volume}{69}},
\bibinfo{pages}{233403} (\bibinfo{year}{2004}).



\bibitem[{\citenamefont{Ewing}(2006)}]{Ewing2006}
\bibinfo{author}{\bibfnamefont{G.~E.} \bibnamefont{Ewing}},
  \bibinfo{journal}{Chem. Rev.} \textbf{\bibinfo{volume}{106}},
  \bibinfo{pages}{1511} (\bibinfo{year}{2006}).



\bibitem[{\citenamefont{F. J. Giessibl
  et~al.}(1992)\citenamefont{Giessibl}}]{Giessibl1992a}
  \bibinfo{author}{\bibfnamefont{F.~J.} \bibnamefont{Giessibl}},
  \bibinfo{journal}{Phys. Rev. B} \textbf{\bibinfo{volume}{45}},
  \bibinfo{pages}{13815} (\bibinfo{year}{1992}).



\bibitem[{\citenamefont{Giessibl} \emph{et~al.}(1999)\citenamefont{Giessibl,
  Bielefeldt, Hembacher, and Mannhart}}]{Giessibl1999a}
\bibinfo{author}{\bibnamefont{Giessibl}, \bibfnamefont{F.~J.}},
  \bibinfo{author}{\bibfnamefont{H.}~\bibnamefont{Bielefeldt}},
  \bibinfo{author}{\bibfnamefont{S.}~\bibnamefont{Hembacher}} \bibnamefont{and}
  \bibinfo{author}{\bibfnamefont{J.}~\bibnamefont{Mannhart}},
  \bibinfo{journal}{{}Appl. Surf. Sci.} \textbf{\bibinfo{volume}{140}},
  \bibinfo{pages}{352} (\bibinfo{year}{1999}).








  \end{thebibliography}
\end{document}